\documentclass[nofootinbib,preprint]{revtex4}%
\usepackage{hyperref}
\usepackage{amsmath}
\usepackage{amsfonts}
\usepackage{amssymb}
\usepackage{graphicx}%
\begin{document}
\title{Higher Spin String States Scattered from D-particle in the Regge Regime and
Factorized Ratios of Fixed Angle Scatterings}
\author{Jen-Chi Lee}
\email{jcclee@cc.nctu.edu.tw}
\affiliation{Department of Electrophysics, National Chiao-Tung University, Hsinchu, Taiwan, R.O.C.}
\author{Yoshihiro Mitsuka}
\email{yoshihiro.mitsuka@gmail.com}
\affiliation{Department of Electrophysics, National Chiao-Tung University, Hsinchu, Taiwan, R.O.C.}
\author{Yi Yang}
\email{yyang@phys.cts.nthu.edu.tw}
\affiliation{Department of Electrophysics, National Chiao-Tung University, Hsinchu, Taiwan, R.O.C.}
\date{\today}

\begin{abstract}
We study scattering of higher spin closed string states at arbitrary mass
levels from D-particle in the Regge regime. We extract the \textit{complete}
infinite ratios among high-energy amplitudes of different string states in the
fixed angle regime from these Regge string scattering amplitudes. In this
calculation, we have used an identity proved recently based on a signless
Stirling number identity in combinatorial theory. The complete ratios
calculated by this indirect method include a subset of ratios calculated
previously by direct fixed angle calculation \cite{Dscatt}. Moreover, we
discover that in spite of the non-factorizability of the closed string
D-particle scattering amplitudes, the complete ratios derived for the fixed
angle regime are found to be factorized. These ratios are consistent with the
decoupling of high-energy zero norm states calculated previously.

\end{abstract}
\maketitle

\bigskip%
\setcounter{equation}{0}
\renewcommand{\theequation}{\arabic{section}.\arabic{equation}}%

\section{Introduction}

Recently high-energy, fixed angle behavior of string scattering amplitudes
\cite{GM, Gross, GrossManes} was intensively investigated for massive
higher-spin string states at arbitrary mass levels \cite{ChanLee1,ChanLee2,
CHL,CHLTY,PRL,paperB,susy,Closed,HL}. The motivation was to uncover the
fundamental hidden stringy spacetime symmetry. An important new ingredient of
this calculation was the zero-norm states (ZNS) \cite{ZNS1,ZNS3,ZNS2} in the
old covariant first quantized string spectrum, in particular, the
identification of inter-particle symmetries induced by the inter-particle ZNS
\cite{ZNS1} in the spectrum. An infinite number of linear relations among
high-energy fixed angle scattering amplitudes of different string states at
each fixed but arbitrary mass levels can be derived. Moreover, these linear
relations can be used to fix the ratios among high-energy scattering
amplitudes of different string states at each fixed mass level. On the other
hand, 2D discrete zero-norm states were also shown \cite{ZNS3} to carry the
spacetime $\omega_{\infty}$ symmetry charges of toy 2D string theory.
Furthermore, in the high-energy limit, these discrete zero-norm states
approach to \cite{PRL,paperB} the discrete Polyakov positive-norm states which
generate the well-known $\omega_{\infty}$ symmetry of the 2D string
\cite{2DString,Winfinity,Ring}. This strongly suggests that the linear
relations obtained from zero-norm states are indeed related to the hidden
symmetry of the 26 dimensional string.

The calculation above was extended to scatterings of bosonic massive closed
string states at arbitrary mass levels from D-brane in \cite{Dscatt,Decay}.
The scattering of massless string states from D-brane was well studied in the
literature and can be found in \cite{Klebanov}. Since the mass of D-brane
scales as the inverse of the string coupling constant $1/g$, it was assumed
that it is infinitely heavy to leading order in $g$ and does not recoil. It
was discovered \cite{Dscatt} that all the scattering amplitudes at arbitrary
energy can be expressed in terms of the generalized hypergeometric function
$_{3}F_{2}$ with special arguments, which terminates to a finite sum and, as a
result, the whole scattering amplitudes consistently reduce to the usual beta
function. For the simple case of D-particle, the authors of \cite{Dscatt}
explicitly calculated high-energy limit of a series of the above scattering
amplitudes for arbitrary mass levels, and derive infinite linear relations
among them for each fixed mass level. The ratios of these high-energy
scattering amplitudes were found to be consistent with the decoupling of
high-energy zero-norm states of the previous works. \cite{ChanLee1,ChanLee2,
CHL,CHLTY,PRL,paperB,susy,Closed}. However, these ratios form only a subset of
the complete ratios for general high-energy vertex in the fixed angle.

In this paper, we calculate the general high-energy scattering amplitudes of
arbitrary higher spin massive closed string states scattered from D-particle
in the small angle or Regge regime (RR). We will assume as before that the
mass of the D-particle is infinitely heavy and so does not recoil. For Regge
string-string scatterings, see \cite{RR1,RR2,RR3,RR4,RR5,RR6}. See also
\cite{OA,DL,KP}. Regge string-string scatterings for arbitrary higher spin
massive states were intensively studied recently in
\cite{bosonic,bosonic2,RRsusy,LYAM,HLY}. In contrast to the case of
scatterings in the fixed angle regime, we will see that there is no linear
relation among string D-particle scatterings in the RR. However, as in the
case of Regge string-string scattering amplitude calculation
\cite{bosonic,bosonic2,RRsusy}, we can extract the infinite fixed angle ratios
of string D-particle scatterings from these Regge string D-particle scattering
amplitudes. In this calculation, we have used a set of identities proved
recently in \cite{LYAM} to extract the fixed angle ratios from the Regge
scattering amplitudes.

We stress that the fixed angle ratios calculated in the present paper by this
indirect method from the Regge calculation are for the most general
high-energy vertex rather than only a subset of ratios \cite{Dscatt} obtained
directly from the fixed angle calculation previously. More importantly, we
discover that the amplitudes calculated in this paper for closed string
D-particle scatterings can not be factorized and thus are different from
amplitudes for the high-energy closed string-string scattering calculated
previously \cite{bosonic2}. Amplitudes for the high-energy closed
string-string scattering can be factorized into two open string scattering
amplitudes by using a calculation \cite{Closed,bosonic2} based on the KLT
formula \cite{KLT}. Presumably, this non-factorization is due to the
non-existence of a KLT-like formula for the string D-brane scattering
amplitudes. There is no physical picture for open string D-particle tree
scattering amplitudes and thus no factorizaion for closed string D-particle
scatterings into two channels of open string D-particle scatterings. However,
we discover that in spite of the non-factorizability of the closed string
D-particle scattering amplitudes, the complete ratios derived for the fixed
angle regime are found to be \textit{factorized}. These ratios are consistent
with the decoupling of high-energy zero norm states calculated previously
\cite{ChanLee1,ChanLee2, CHL,CHLTY,PRL,paperB,susy,Closed}.

\textit{ }This paper is organized as follows. In section II, we first set up
the kinematics. In section III, we calculate the general string D-particle
scatterings in the RR. In section IV, we extract the ratios of string
D-particle fixed angle scattering amplitudes from RR amplitudes. We also
discuss and compare the ratios of string D-particle and string-string
scatterings. Finally, we give a brief conclusion in section V.%

\setcounter{equation}{0}
\renewcommand{\theequation}{\arabic{section}.\arabic{equation}}%

\section{Kinematics Set-up}

In this paper, we consider an incoming string state with momentum $k_{2}$
scattered from an infinitely heavy D-particle and end up with string state
with momentum $k_{1}$in the RR. The high-energy scattering plane will be
assumed to be the $X-Y$ plane, and the momenta are arranged to be%
\begin{align}
k_{1}  &  =\left(  E,\mathrm{k}_{1}\cos\phi,-\mathrm{k}_{1}\sin\phi\right)
,\\
k_{2}  &  =\left(  -E,-\mathrm{k}_{2},0\right)
\end{align}
where%
\begin{equation}
E=\sqrt{\mathrm{k}_{2}^{2}+M_{2}^{2}}=\sqrt{\mathrm{k}_{1}^{2}+M_{1}^{2}},
\label{II3}%
\end{equation}
and $\phi$ is the scattering angle. For simplicity, we will calculate the disk
amplitude in this paper. The relevant propagators for the left-moving string
coordinate $X^{\mu}\left(  z\right)  $ and the right-moving one $\tilde
{X}^{\nu}\left(  \bar{w}\right)  $ are%
\begin{align}
\left\langle X^{\mu}\left(  z\right)  ,X^{\nu}\left(  w\right)  \right\rangle
&  =-\eta^{\mu\nu}\left\langle X\left(  z\right)  ,X\left(  w\right)
\right\rangle =-\eta^{\mu\nu}\ln\left(  z-w\right)  ,\label{DD1}\\
\left\langle \tilde{X}^{\mu}\left(  \bar{z}\right)  ,\tilde{X}^{\nu}\left(
\bar{w}\right)  \right\rangle  &  =-\eta^{\mu\nu}\left\langle \tilde{X}\left(
\bar{z}\right)  ,\tilde{X}\left(  \bar{w}\right)  \right\rangle =-\eta^{\mu
\nu}\ln\left(  \bar{z}-\bar{w}\right)  ,\label{DD2}\\
\left\langle X^{\mu}\left(  z\right)  ,\tilde{X}^{\nu}\left(  \bar{w}\right)
\right\rangle  &  =-D^{\mu\nu}\left\langle X\left(  z\right)  ,\tilde
{X}\left(  \bar{w}\right)  \right\rangle =-D^{\mu\nu}\ln\left(  1-z\bar
{w}\right)  \text{\ \ (for Disk)} \label{DD}%
\end{align}
where matrix $D$ has the standard form for the fields satisfying Neumann
boundary condition, while $D$ reverses the sign for the fields satisfying
Dirichlet boundary condition. Instead of the Mandelstam variables used in the
string-string scatterings, we define%
\begin{align}
a_{0}  &  \equiv k_{1}\cdot D\cdot k_{1}=-E^{2}-\mathrm{k}_{1}^{2}\sim
-2E^{2},\\
a_{0}^{\prime}  &  \equiv k_{2}\cdot D\cdot k_{2}=-E^{2}-\mathrm{k}_{2}%
^{2}\sim-2E^{2},\\
b_{0}  &  \equiv2k_{1}\cdot k_{2}+1=2\left(  E^{2}-\mathrm{k}_{1}%
\mathrm{k}_{2}\cos\phi\right)  +1=fixed,\label{II4}\\
c_{0}  &  \equiv2k_{1}\cdot D\cdot k_{2}+1=2\left(  E^{2}+\mathrm{k}%
_{1}\mathrm{k}_{2}\cos\phi\right)  +1,
\end{align}
so that%
\begin{equation}
2a_{0}+b_{0}+c_{0}=2M_{1}^{2}+2.
\end{equation}
Since we are going to calculate Regge scattering amplitudes, $b_{0}=fixed$. We
can use Eq.(\ref{II3}) and Eq.(\ref{II4}) to calculate%
\begin{align}
\cos\phi\sim &  1-\frac{b_{0}-M_{1}^{2}-M_{2}^{2}-1}{2\mathrm{k}_{1}^{2}}\\
\sin\phi\sim &  \frac{\sqrt{b_{0}-M_{1}^{2}-M_{2}^{2}-1}}{\mathrm{k}_{1}%
}\equiv\frac{\sqrt{\tilde{b}_{0}}}{\mathrm{k}_{1}}%
\end{align}
The normalized polarization vectors on the high-energy scattering plane of the
$k_{2}$ string state are defined to be \cite{ChanLee1,ChanLee2}
\begin{equation}
e_{P}=\frac{1}{M_{2}}(-E,-\mathrm{k}_{2},0)=\frac{k_{2}}{M_{2}},
\end{equation}%
\begin{equation}
e_{L}=\frac{1}{M_{2}}(-\mathrm{k}_{2},-E,0),
\end{equation}%
\begin{equation}
e_{T}=(0,0,1).
\end{equation}
One can then easily calculate the following kinematics%
\begin{align}
e^{T}\cdot k_{2}  &  =0,\nonumber\\
e^{T}\cdot k_{1}  &  =-\mathrm{k}_{1}\sin\phi\sim-\sqrt{\tilde{b}_{0}%
},\nonumber\\
e^{T}\cdot D\cdot k_{1}  &  =\mathrm{k}_{1}\sin\phi\sim\sqrt{\tilde{b}_{0}%
},\nonumber\\
e^{T}\cdot D\cdot k_{2}  &  =0,\nonumber\\
e^{P}\cdot k_{2}  &  =-M_{2},\nonumber\\
e^{P}\cdot k_{1}  &  =\frac{1}{M_{2}}\left[  E^{2}-\mathrm{k}_{1}%
\mathrm{k}_{2}\cos\phi\right]  =\frac{b_{0}-1}{2M_{2}},\nonumber\\
e^{P}\cdot D\cdot k_{1}  &  =\frac{1}{M_{2}}\left[  E^{2}+\mathrm{k}%
_{1}\mathrm{k}_{2}\cos\phi\right]  =\frac{c_{0}-1}{2M_{2}},\nonumber\\
e^{P}\cdot D\cdot k_{2}  &  =\frac{1}{M_{2}}\left[  -E^{2}-\mathrm{k}_{2}%
^{2}\right]  =\frac{a^{\prime}_{0}}{M_{2}}\sim\frac{a_{0}}{M_{2}},\nonumber\\
e^{T}\cdot D\cdot e^{T}  &  =-1,\nonumber\\
e^{T}\cdot D\cdot e^{P}  &  =e^{P}\cdot D\cdot e^{T}=0,\nonumber\\
e^{P}\cdot D\cdot e^{P}  &  =\frac{1}{M_{2}^{2}}\left[  -E^{2}-\mathrm{k}%
_{2}^{2}\right]  =\frac{a^{\prime}_{0}}{M_{2}^{2}}\sim\frac{a_{0}}{M_{2}^{2}},
\label{Kine}%
\end{align}
which will be useful in the amplitude calculation in the next section.%

\setcounter{equation}{0}
\renewcommand{\theequation}{\arabic{section}.\arabic{equation}}%

\section{Regge String D-particle Scatterings}

\bigskip We now begin to calculate the scattering amplitudes. For simplicity,
we will take $k_{1}$ to be the tachyon and $k_{2}$ to be the tensor states.
One can easily argue that a class of high-energy string states for $k_{2}$ in
the RR are \cite{bosonic,RRsusy}%
\begin{equation}
|p_{n},p_{n}^{\prime},q_{m},q_{m}^{\prime}\rangle=\left[  \prod_{n>0}\left(
\alpha_{-n}^{T}\right)  ^{p_{n}}\prod_{m>0}\left(  \alpha_{-m}^{P}\right)
^{q_{m}}\right]  \left[  \prod_{n>0}\left(  \tilde{\alpha}_{-n}^{T}\right)
^{p_{n}^{\prime}}\prod_{m>0}\left(  \tilde{\alpha}_{-m}^{P}\right)
^{q_{m}^{\prime}}\right]  |0,k\rangle\label{general states}%
\end{equation}
with%
\begin{align}
\sum_{n}n\left(  p_{n}-p_{n}^{\prime}\right)  +\sum_{m}m\left(  q_{m}%
-q_{m}^{\prime}\right)   &  =0,\\
\sum_{n}n\left(  p_{n}+p_{n}^{\prime}\right)  +\sum_{m}m\left(  q_{m}%
+q_{m}^{\prime}\right)   &  =N=\text{const}%
\end{align}
where $M_{2}^{2}=(N-2).$

\subsection{An example}

Before calculating the string D-particle scattering amplitudes for general
cases, we take an example and illustrate the method of calculation. We
consider the case
\begin{align}
&  p_{1}=p^{\prime}_{1}=q_{1}=q^{\prime}_{1}=q_{2}=q^{\prime}_{2}%
=1,\quad\mathrm{others}=0.\label{example-state}%
\end{align}
As we will see in the next subsection, the string D-particle scattering
amplitudes with the general states (\ref{general states}) are reduced to
simple forms in the Regge limit, in which most of the ways of contracting the
operators are discarded as subleading. For a fixed number of the contractions
between $\partial X^{P}$ and $\bar{\partial} \tilde X^{P}$, the ways of
contracting the other factors are determined by the following rules.
\begin{align}
\alpha_{-n}^{T}  & \quad\text{1 term (contraction of $i k_{1} X$ with
$\partial_{n} X^{T}$)}\label{rule1}\\
\tilde{\alpha}_{-n}^{T}  & \quad\text{1 term (contraction of $i k_{1}
\tilde{X}$ with $\bar{\partial}_{n} \tilde{X}^{T}$)}\label{rule2}\\
\alpha_{-n}^{P}  & \quad%
\begin{cases}
\left(  n>1 \right)  \quad\text{1 term (contraction of $i k_{1} X$ with
$\partial_{n} X^{P}$)}\\
\left(  n=1 \right)  \quad\text{2 terms (contraction of $i k_{1} X$ and $i
k_{2} X$ with $\partial X^{P}$)}%
\end{cases}
\label{rule3}\\
\tilde{\alpha}_{-n}^{P}  & \quad%
\begin{cases}
\left(  n>1 \right)  \quad\text{1 term (contraction of $i k_{1} \tilde{X}$
with $\bar{\partial}_{n} \tilde{X}^{P}$ )}\\
\left(  n=1 \right)  \quad\text{2 terms (contraction of $i k_{1} \tilde{X}$
and $i k_{2} \tilde{X}$ with $\bar{\partial} \tilde{X}^{P}$ )}%
\end{cases}
\label{rule4}%
\end{align}
Therefore we take the state Eq.(\ref{example-state}) as the simplest example
for the purpose of this subsection.

We start with the procedure in \cite{KLT} to treat the vertex operator
corresponding to the state (\ref{example-state}).
\begin{align}
V  & =i^{6}\varepsilon_{\mu_{1}\cdots\mu_{6}}:\partial X^{\mu_{1}}\partial
X^{\mu_{2}} \partial^{2} X^{\mu_{3}}e^{ik_{2} X}\left(  z\right)
:\ :\bar{\partial} \tilde{X}^{\mu_{4}}\bar{\partial} \tilde{X}^{\mu_{5}}
\bar{\partial}^{2} \tilde{X}^{\mu_{6}}e^{ik_{2} \tilde{X}}\left(  \bar
{z}\right) :\nonumber\\
& =i^{6}:\partial X^{T}\partial X^{P} \partial^{2} X^{P}e^{ik_{2} X}\left(
z\right) :\ :\bar{\partial} \tilde{X}^{T}\bar{\partial} \tilde{X}^{P}
\bar{\partial}^{2} \tilde{X}^{P}e^{ik_{2} \tilde{X}}\left(  \bar{z}\right)
:\nonumber\\
& = i^{6} \left[  :\exp\left\{  ik_{2} X(z) + \varepsilon_{T}^{(1)}\partial
X^{T} (z) + \varepsilon_{P}^{(1)}\partial X^{P} (z) + \varepsilon_{P}%
^{(2)}\partial^{2} X ^{P} (z) \right\} : \right. \nonumber\\
&  \quad\quad\times\left.  :\exp\left\{  ik_{2} \tilde{X}(\bar{z}) +
\varepsilon_{T}^{\prime(1)}\partial\tilde{X}^{T} (\bar{z}) + \varepsilon
_{P}^{\prime(1)}\partial\tilde{X}^{P} (\bar{z}) + \varepsilon_{P}^{\prime
(2)}\partial^{2} \tilde{X}^{P} (\bar{z}) \right\} : \right]
_{\mathrm{linear\ terms}}\label{exponentiation-ex}%
\end{align}
In the last equation, we have introduced the dummy variables $\varepsilon
_{T}^{(1)}, \varepsilon_{P}^{(1)}, \varepsilon_{P}^{(2)}, \varepsilon
_{T}^{\prime(1)},\varepsilon_{P}^{\prime(1)},\varepsilon_{P}^{\prime(2)}$
associated with the non-vanishing component $\varepsilon_{TPPTPP}$ of the
polarization tensor and written the operator in the exponential form. ``linear
terms'' indicate that we take the sum of the terms linear in all of
$\varepsilon_{T}^{(1)},\varepsilon_{P}^{(1)}, \varepsilon_{P}^{(2)},
\varepsilon_{T}^{\prime(1)},\varepsilon_{P}^{\prime(1)}$, and $\varepsilon
_{P}^{\prime(2)}$. This sum can be rephrased as the coefficient of the product
$\varepsilon_{T}^{(1)} \varepsilon_{P}^{(1)} \varepsilon_{P}^{(2)}
\varepsilon_{T}^{\prime(1)}\varepsilon_{P}^{\prime(1)} \varepsilon_{P}%
^{\prime(2)}$ because we set the dummy variables to be 1 at the end of calculation.

The string D-particle scattering amplitudes can be calculated to be
\begin{align}
A  &  =\int d^{2} z_{1} d^{2} z_{2}\ i^{6}\nonumber\\
& \cdot\left\langle :e^{ik_{1}X}\left(  z_{1}\right) : : e^{ik_{1}\tilde{X}%
}\left(  \bar{z}_{1}\right) : :\partial X^{T}\partial X^{P} \partial^{2}
X^{P}e^{ik_{2} X}\left(  z_{2} \right) : :\bar{\partial} \tilde{X}^{T}%
\bar{\partial} \tilde{X}^{P} \bar{\partial}^{2} \tilde{X}^{P}e^{ik_{2}
\tilde{X}}\left(  \bar{z}_{2}\right) : \right\rangle \label{def-amp}\\
&  =i^{6}\int d^{2}z_{1}d^{2}z_{2}\nonumber\\
& \cdot\left[  \exp\left\{  \left\langle ik_{1}X \left(  z_{1}\right)
\ ik_{1}\tilde{X} \left(  \bar{z}_{1}\right)  \right\rangle \right.  \right.
\nonumber\\
&  +\left\langle \left(  \varepsilon^{(1)}_{T}\partial X^{T} +\varepsilon
^{(1)}_{P}\partial X^{P} +\varepsilon^{(2)}_{P}\partial^{2} X^{P}+ik_{2}X
\right)  \left(  z_{2}\right)  \right. \nonumber\\
&  \left. \qquad\quad\cdot\left(  \varepsilon_{T}^{\prime(1)}\bar{\partial
}\tilde{X}^{T} +\varepsilon_{P}^{\prime(1)}\bar{\partial}\tilde{X}^{P}
+\varepsilon_{P}^{\prime(2)}\bar{\partial}^{2} \tilde{X}^{P} +ik_{2}\tilde{X}
\right)  \left(  \bar{z}_{2}\right)  \right\rangle \nonumber\\
&  +\left\langle ik_{1}X \left(  z_{1}\right)  \left(  \varepsilon^{(1)}%
_{T}\partial X^{T} +\varepsilon_{P}^{(1)} \partial X^{P} +\varepsilon
_{P}^{(2)} \partial^{2} X^{P}+ik_{2}X \right)  \left(  z_{2}\right)
\right\rangle \nonumber\\
&  +\left\langle ik_{1}\tilde{X} \left(  \bar{z}_{1}\right)  \left(
\varepsilon_{T}^{\prime(1)} \bar{\partial} \tilde{X}^{T} +\varepsilon
_{P}^{\prime(1)}\bar{\partial}\tilde{X}^{P} +\varepsilon_{P}^{\prime(2)}%
\bar{\partial}^{2}\tilde{X}^{P} +ik_{2}\tilde{X} \right)  \left(  \bar{z}%
_{2}\right)  \right\rangle \nonumber\\
&  +\left\langle ik_{1}X \left(  z_{1}\right)  \left(  \varepsilon_{T}%
^{\prime(1)}\bar{\partial}\tilde{X}^{T} +\varepsilon_{P}^{\prime(1)}%
\bar{\partial}\tilde{X}^{P} +\varepsilon_{P}^{\prime(2)}\bar{\partial}%
^{2}\tilde{X}^{P} +ik_{2}\tilde{X} \right)  \left(  \bar{z}_{2} \right)
\right\rangle \nonumber\\
&  \left.  \left.  +\left\langle ik_{1}\tilde{X} \left(  \bar{z}_{1}\right)
\left(  \varepsilon_{T}^{(1)}\partial X^{T} +\varepsilon_{P}^{(1)}\partial
X^{P} +\varepsilon_{P}^{(2)}\partial^{2} X^{P} +ik_{2}X\right)  \left(
z_{2}\right)  \right\rangle \right\}  \right] _{\text{linear terms}}\nonumber
\end{align}
\begin{align}
&  =\int d^{2}z_{1}d^{2}z_{2} \left\langle : e^{ik_{1}X}\left(  z_{1}\right)
: : e^{ik_{1}\tilde{X}}\left(  \bar{z}_{1}\right)  : : e^{ik_{2} X}\left(
z_{2}\right)  : : e^{ik_{2}\tilde{X}}\left(  \bar{z}_{2}\right)  :
\right\rangle \nonumber\\
&  \cdot\Big[\exp\Big\{\nonumber\\
&  -\varepsilon_{T}^{(1)} \left[  ie^{T} k_{1}\partial_{2} \left\langle
X\left(  z_{1}\right)  X\left(  z_{2}\right)  \right\rangle +ie^{T} D
k_{1}\partial_{2} \left\langle \tilde{X}\left(  \bar{z}_{1}\right)  X\left(
z_{2}\right)  \right\rangle +ie^{T} D k_{2}\partial_{2}\left\langle \tilde
{X}\left(  \bar{z}_{2}\right)  X\left(  z_{2}\right)  \right\rangle \right]
\nonumber\\
&  -\varepsilon_{T}^{\prime(1)} \left[  ie^{T} D k_{1}\bar{\partial}_{2}
\left\langle X\left(  z_{1}\right)  \tilde{X}\left(  \bar{z}_{2}\right)
\right\rangle +ie^{T} k_{1}\bar{\partial}_{2} \left\langle \tilde{X}\left(
\bar{z}_{1}\right)  \tilde{X}\left(  \bar{z}_{2}\right)  \right\rangle +ie^{T}
D k_{2}\bar{\partial}_{2} \left\langle X\left(  z_{2}\right)  \tilde{X}\left(
\bar{z}_{2}\right)  \right\rangle \right] \nonumber\\
&  -\varepsilon_{P}^{(1)} \left[  ie^{P} k_{1}\partial_{2} \left\langle
X\left(  z_{1}\right)  X\left(  z_{2}\right)  \right\rangle +ie^{P} D
k_{1}\partial_{2} \left\langle \tilde{X}\left(  \bar{z}_{1}\right)  X\left(
z_{2}\right) \right\rangle +ie^{P} D k_{2}\partial_{2} \left\langle \tilde
{X}\left(  \bar{z}_{2}\right)  X\left(  z_{2}\right)  \right\rangle \right]
\nonumber\\
&  -\varepsilon_{P}^{(2)} \left[  ie^{P} k_{1}\partial_{2}^{2} \left\langle
X\left(  z_{1}\right)  X\left(  z_{2}\right)  \right\rangle +ie^{P} D
k_{1}\partial_{2}^{2} \left\langle \tilde{X}\left(  \bar{z}_{1}\right)
X\left(  z_{2}\right) \right\rangle +ie^{P} D k_{2}\partial_{2}^{2}
\left\langle \tilde{X}\left(  \bar{z}_{2}\right)  X\left(  z_{2}\right)
\right\rangle \right] \nonumber\\
&  -\varepsilon_{P}^{\prime(1)} \left[  ie^{P} D k_{1}\bar{\partial}_{2}
\left\langle X\left(  z_{1}\right)  \tilde{X}\left(  \bar{z}_{2}\right)
\right\rangle +ie^{P} k_{1}\bar{\partial}_{2} \left\langle \tilde{X}\left(
\bar{z}_{1}\right)  \tilde{X}\left(  \bar{z}_{2}\right)  \right\rangle +ie^{P}
D k_{2}\bar{\partial}_{2} \left\langle X\left(  z_{2}\right)  \tilde{X}\left(
\bar{z}_{2}\right)  \right\rangle \right] \nonumber\\
&  -\varepsilon_{P}^{\prime(2)} \left[  ie^{P} D k_{1}\bar{\partial}_{2}^{2}
\left\langle X\left(  z_{1}\right)  \tilde{X}\left(  \bar{z}_{2}\right)
\right\rangle +ie^{P} k_{1}\bar{\partial}_{2}^{2} \left\langle \tilde
{X}\left(  \bar{z}_{1}\right)  \tilde{X}\left(  \bar{z}_{2}\right)
\right\rangle +ie^{P} D k_{2}\bar{\partial}_{2}^{2} \left\langle X\left(
z_{2}\right)  \tilde{X}\left(  \bar{z}_{2}\right)  \right\rangle \right]
\nonumber\\
&  -\varepsilon_{T}^{(1)}\varepsilon_{T}^{\prime(1)} \left[  e^{T} D e^{T}
\partial\bar{\partial} \left\langle X\left(  z_{2}\right) \tilde{X}\left(
\bar{z}_{2}\right)  \right\rangle \right] \nonumber\\
& -\varepsilon_{P}^{(1)}\varepsilon_{P}^{\prime(1)} \left[  e^{P} D e^{P}
\partial\bar{\partial} \left\langle X\left(  z_{2}\right) \tilde{X}\left(
\bar{z}_{2}\right)  \right\rangle \right]  -\varepsilon_{P}^{(1)}%
\varepsilon_{P}^{\prime(2)} \left[  e^{P} D e^{P} \partial\bar{\partial}^{2}
\left\langle X\left(  z_{2}\right) \tilde{X}\left(  \bar{z}_{2}\right)
\right\rangle \right] \nonumber\\
& -\varepsilon_{P}^{(2)}\varepsilon_{P}^{\prime(1)} \left[  e^{P} D e^{P}
\partial^{2}\bar{\partial} \left\langle X\left(  z_{2}\right) \tilde{X}\left(
\bar{z}_{2}\right)  \right\rangle \right]  -\varepsilon_{P}^{(2)}%
\varepsilon_{P}^{\prime(2)} \left[  e^{P} D e^{P} \partial^{2} \bar{\partial
}^{2} \left\langle X\left(  z_{2}\right) \tilde{X}\left(  \bar{z}_{2}\right)
\right\rangle \right] \nonumber\\
&  -\varepsilon_{T}^{(1)}\varepsilon_{P}^{\prime(1)} \left[  e^{T} D e^{P}
\partial\bar{\partial} \left\langle X\left(  z_{2}\right)  \tilde{X}\left(
\bar{z}_{2}\right)  \right\rangle \right]  -\varepsilon_{T}^{(1)}%
\varepsilon_{P}^{\prime(2)} \left[  e^{T} D e^{P} \partial\bar{\partial}^{2}
\left\langle X\left(  z_{2}\right)  \tilde{X}\left(  \bar{z}_{2}\right)
\right\rangle \right] \nonumber\\
&  -\varepsilon_{P}^{(1)}\varepsilon_{T}^{\prime\ (1)} \left[  e^{P} D e^{T}
\partial\bar{\partial} \left\langle X\left(  z_{2}\right)  \tilde{X}\left(
\bar{z}_{2}\right)  \right\rangle \right]  -\varepsilon_{P}^{(2)}%
\varepsilon_{T}^{\prime\ (1)} \left[  e^{P} D e^{T} \partial^{2}\bar{\partial}
\left\langle X\left(  z_{2}\right)  \tilde{X}\left(  \bar{z}_{2}\right)
\right\rangle \right]  \Big\}\Big]_{\text{linear terms}}\nonumber
\end{align}
\begin{align}
&  =\int d^{2}z_{1}d^{2}z_{2} \left(  1-z_{1}\bar{z}_{1}\right) ^{a_{0}%
}\left(  1-z_{2}\bar{z}_{2}\right) ^{a_{0}^{\prime}} \left\vert z_{1}%
-z_{2}\right\vert ^{b_{0}-1}\left\vert 1-z_{1}\bar{z}_{2}\right\vert
^{c_{0}-1}\nonumber\\
& \cdot\Big[\exp\Big\{\nonumber\\
&  \varepsilon_{T}^{(1)} \left[  \dfrac{ie^{T} k_{1}}{\left(  z_{1}%
-z_{2}\right) } +\dfrac{i e^{T} D k_{1} \bar{z}_{1}}{\left( 1-\bar{z}_{1}%
z_{2}\right) } +\dfrac{i e^{T} D k_{2} \bar{z}_{2}}{\left( 1-\bar{z}_{2}%
z_{2}\right) } \right]  +\varepsilon_{T}^{\prime(1)} \left[  \dfrac{i e^{T} D
k_{1} z_{1}}{\left(  1-z_{1}\bar{z}_{2}\right)  } +\dfrac{i e^{T} k_{1}%
}{\left(  \bar{z}_{1}-\bar{z}_{2}\right)  } +\dfrac{i e^{T} D k_{2} z_{2}%
}{\left(  1-z_{2}\bar{z}_{2}\right) } \right] \nonumber\\
&  +\varepsilon_{P}^{(1)} \left[  \dfrac{i e^{P} k_{1}}{\left(  z_{1}%
-z_{2}\right) } +\dfrac{i e^{P} D k_{1} \bar{z}_{1}}{\left(  1-\bar{z}%
_{1}z_{2}\right) } +\dfrac{i e^{P} D k_{2} \bar{z}_{2}}{\left(  1-\bar{z}%
_{2}z_{2}\right) } \right]  +\varepsilon_{P}^{(2)} \left[  \dfrac{i e^{P}
k_{1}}{\left(  z_{1}-z_{2}\right) ^{2}} +\dfrac{i e^{P} D k_{1} \bar{z}%
_{1}^{2}}{\left(  1-\bar{z}_{1}z_{2}\right) ^{2}} +\dfrac{i e^{P} D k_{2}
\bar{z}_{2}^{2}}{\left(  1-\bar{z}_{2}z_{2}\right) ^{2}} \right] \nonumber\\
&  +\varepsilon_{P}^{\prime(1)} \left[  \dfrac{i e^{P} D k_{1} z_{1}}{\left(
1-z_{1}\bar{z}_{2}\right) } +\dfrac{i e^{P} k_{1}}{\left(  \bar{z}_{1}-\bar
{z}_{2}\right) } +\dfrac{i e^{P} D k_{2} z_{2}}{\left(  1-z_{2}\bar{z}%
_{2}\right)  } \right]  +\varepsilon_{P}^{\prime(2)} \left[  \dfrac{i e^{P} D
k_{1} z_{1}^{2}}{\left(  1-z_{1}\bar{z}_{2}\right) ^{2}} +\dfrac{i e^{P}
k_{1}}{\left(  \bar{z}_{1}-\bar{z}_{2}\right) ^{2}} +\dfrac{i e^{P} D k_{2}
z_{2}^{2}}{\left(  1-z_{2}\bar{z}_{2}\right) ^{2} } \right] \nonumber\\
&  +\varepsilon_{T}^{(1)}\varepsilon_{T}^{\prime(1)} \dfrac{ e^{T} D e^{T}
}{\left(  1-z_{2} \bar{z}_{2}\right) ^{2}}\nonumber\\
&  +\varepsilon_{P}^{(1)}\varepsilon_{P}^{\prime(1)} \dfrac{ e^{P} D e^{P}
}{\left(  1-z_{2} \bar{z}_{2}\right) ^{2}} +2\varepsilon_{P}^{(1)}%
\varepsilon_{P}^{\prime(2)} \dfrac{ e^{P} D e^{P} z_{2}}{\left(  1-z_{2}
\bar{z}_{2} \right) ^{3}} +2\varepsilon_{P}^{(2)}\varepsilon_{P}^{\prime(1)}
\dfrac{ e^{P} D e^{P} \bar{z}_{2}}{\left(  1-z_{2} \bar{z}_{2}\right) ^{3}}
+2\varepsilon_{P}^{(2)}\varepsilon_{P}^{\prime(2)} \dfrac{ e^{P} D e^{P}
\left(  1+2z_{2} \bar{z}_{2} \right) }{\left(  1-z_{2} \bar{z}_{2}\right)
^{4}}\nonumber\\
&  +\varepsilon_{T}^{(1)}\varepsilon_{P}^{\prime(1)} \dfrac{ e^{T} D e^{P}%
}{\left(  1-z_{2} \bar{z}_{2}\right) ^{2}} +2\varepsilon_{T}^{(1)}%
\varepsilon_{P}^{\prime(2)} \dfrac{ e^{T} D e^{P} z_{2}}{\left(  1-z_{2}
\bar{z}_{2}\right) ^{3}} +\varepsilon_{P}^{(1)}\varepsilon_{T}^{\prime(1)}
\dfrac{ e^{P} D e^{T}}{\left(  1-z_{2} \bar{z}_{2}\right) ^{2}} +2\varepsilon
_{P}^{(2)}\varepsilon_{T}^{\prime(1)} \dfrac{ e^{P} D e^{T} \bar{z}_{2}%
}{\left(  1-z_{2} \bar{z}_{2}\right) ^{3}} \nonumber\\
&  \Big\}\Big]_{\text{linear terms}}%
\end{align}
To fix the $SL(2,R)$ modulus group on the disk, we set $z_{1}=0$ and $z_{2}%
=r$, then $d^{2}z_{1}d^{2}z_{2}=d\left(  r^{2}\right)  .$ By using
Eq.(\ref{Kine}), the amplitude can then be reduced to%
\begin{align}
A  &  =\int_{0}^{1} d\left(  r^{2}\right)  \left(  1-r^{2}\right)
^{a_{0}^{\prime}}r^{b_{0}-1}\nonumber\\
& \cdot\Big[\nonumber\\
&  \exp\left\{
\begin{array}
[c]{l}%
\varepsilon_{T}^{(1)} \left[  -\dfrac{i\sqrt{\tilde{b}_{0}}}{-r}\right]
+\varepsilon_{T}^{\prime(1)} \left[  -\dfrac{i\sqrt{\tilde{b}_{0}}}{-r}\right]
\\
+\varepsilon_{P}^{(1)} \left[  \dfrac{i\frac{b_{0}-1}{2M_{2}}}{-r}
+\dfrac{i\frac{a_{0}}{M_{2}}}{ \left(  1-r^{2}\right)  /r } \right]
+\varepsilon_{P}^{(2)} \left[  \dfrac{i\frac{b_{0}-1}{2M_{2}}}{\left(
-r\right) ^{2}} +\dfrac{i\frac{a_{0}}{M_{2}}}{ \left[  \left(  1-r^{2}\right)
/r \right] ^{2}} \right] \\
+\varepsilon_{P}^{\prime(1)} \left[  \dfrac{i\frac{b_{0}-1}{2M_{2}}}{-r}
+\dfrac{i\frac{a_{0}}{M_{2}}}{ \left(  1-r^{2}\right) /r} \right]
+\varepsilon_{P}^{\prime(2)} \left[  \dfrac{i\frac{b_{0}-1}{2M_{2}}}{\left(
-r \right) ^{2}} +\dfrac{i\frac{a_{0}}{M_{2}}}{ \left[ 1-r^{2}/r\right] ^{2}}
\right] \\
-\varepsilon_{T}^{(1)}\varepsilon_{T}^{\prime(1)}\dfrac{1}{\left(
1-r^{2}\right) ^{2}}\\
+\varepsilon_{P}^{(1)}\varepsilon_{P}^{\prime(1)} \dfrac{\frac{a_{0}}%
{M_{2}^{2}}}{\left(  1-r^{2}\right) ^{2}} +2\varepsilon_{P}^{(1)}%
\varepsilon_{P}^{\prime(2)} \dfrac{\frac{a_{0}}{M_{2}^{2}}r}{\left(
1-r^{2}\right) ^{3}} +2\varepsilon_{P}^{(2)}\varepsilon_{P}^{\prime(1)}
\dfrac{\frac{a_{0}}{M_{2}^{2}}r}{\left(  1-r^{2}\right) ^{3}} +2\varepsilon
_{P}^{(2)}\varepsilon_{P}^{\prime(2)} \dfrac{\frac{a_{0}}{M_{2}^{2}} \left(
1+2r^{2}\right) }{\left(  1-r^{2}\right) ^{4}}%
\end{array}
\right\} \nonumber\\
& \Big]_{\text{linear terms}}\label{ex1}%
\end{align}

Although in Eq.(\ref{ex1}) we have dropped several subleading terms by using
the kinematic relations Eq.(\ref{Kine}), Eq.(\ref{ex1}) still has subleading
terms. We can see that by performing the integration of a generic term in
Eq.(\ref{ex1}) and looking at its behavior in the Regge limit explicitly.
\begin{align}
\int_{0}^{1} d\left(  r^{2}\right)  \left(  1-r^{2}\right) ^{a_{0}^{\prime
}+n_{a}}r^{b_{0}-1-N+n_{b}}  & =B\left(  a^{\prime}_{0} +1 +n_{a}, \frac{b_{0}
-N+1}{2}+\frac{n_{b}}{2}\right) \nonumber\\
& =B\left(  a^{\prime}_{0} +1 , \frac{b_{0} -N+1}{2} \right)  \frac{\left(
a^{\prime}_{0} +1\right) _{n_{a}} \left(  \frac{b_{0} -N+1}{2}\right)
_{\frac{n_{b}}{2}}} {\left(  a^{\prime}_{0} +1 + \frac{b_{0} -N+1}{2}\right)
_{n_{a} +\frac{n_{b}}{2}}}\nonumber\\
& \sim B\left(  a_{0} +1 , \frac{b_{0} -N+1}{2} \right)  \left(  \frac{b_{0}
-N+1}{2}\right) _{\frac{n_{b}}{2}} \left(  a_{0} \right) ^{-\frac{n_{b}}{2}%
}\label{Regge behavior}%
\end{align}
Here the Pochhammer symbol is defined by $\left(  x\right) _{y}=\frac
{\Gamma\left( x+y\right) }{\Gamma\left( x\right) }$ , which, if $y$ is a
positive integer, is reduced to $\left(  x\right) _{y} = x (x+1)(x+2)\cdots
(x+y-1). $ From the Regge behavior Eq.(\ref{Regge behavior}), we see that
increasing one power of $1/r$ in the integrand results in increasing one-half
power of $a_{0}$. Thus we obtain the following rules to determine which terms
in the exponent of Eq.(\ref{ex1}) contribute to the leading behavior of the
amplitude:
\begin{equation}
1/r\rightarrow E,\quad a_{0}\rightarrow E^{2}.\label{rule5}%
\end{equation}

We can now drop the subleading terms in energy to get%
\begin{align}
A  &  =\int_{0}^{1} d\left(  r^{2}\right)  \left(  1-r^{2}\right)
^{a_{0}^{\prime}}r^{b_{0}-1}\nonumber\\
&  \cdot\left[  \exp\left\{  \varepsilon_{T}^{(1)} \left[  -\dfrac
{i\sqrt{\tilde{b}_{0}}}{-r}\right]  +\varepsilon_{T}^{\prime(1)} \left[
-\dfrac{i\sqrt{\tilde{b}_{0}}}{-r}\right]  +\varepsilon_{P}^{(2)} \left[
\dfrac{i\frac{b_{0}-1}{2M_{2}}}{\left( -r\right) ^{2}} \right]  +\varepsilon
_{P}^{\prime(2)} \left[  \dfrac{i\frac{b_{0}-1}{2M_{2}}}{\left(  -r \right)
^{2}} \right]  \right\}  \right] _{\epsilon_{TPTP}}\nonumber\\
&  \cdot\left[  \exp\left\{  \varepsilon_{P}^{(1)} \left[  \dfrac{i\frac
{b_{0}-1}{2M_{2}}}{-r} +\dfrac{i\frac{a_{0}}{M_{2}}}{ \left(  1-r^{2}\right)
/r } \right]  +\varepsilon_{P}^{\prime(1)} \left[  \dfrac{i\frac{b_{0}%
-1}{2M_{2}}}{-r} +\dfrac{i\frac{a_{0}}{M_{2}}}{ \left(  1-r^{2}\right) /r}
\right]  +\varepsilon_{P}^{(1)}\varepsilon_{P}^{\prime(1)} \dfrac{\frac{a_{0}%
}{M_{2}^{2}}}{\left(  1-r^{2}\right) ^{2}} \right\}  \right] _{\epsilon_{PP}%
}\label{ex2}%
\end{align}
where $[\cdots]_{\epsilon_{TPTP}}$ in the second line and $[\cdots
]_{\epsilon_{PP}}$ in the third line indicate that we take the coefficients of
$\varepsilon_{T}^{(1)}\varepsilon_{T}^{\prime(1)}\varepsilon_{P}%
^{(2)}\varepsilon_{P}^{\prime(2)}$ and $\varepsilon_{P}^{(1)}\varepsilon
_{P}^{\prime(1)}$ respectively. Because of the difference in the powers of
$1/r$ and $a_{0}$ in the exponent of Eq.(\ref{ex1}), Eq.(\ref{ex2}) has much
more structure for $\varepsilon_{P}^{(1)}$ and $\varepsilon_{P}^{\prime(1)}$
than for $\varepsilon_{T}^{(1)}$, $\varepsilon_{T}^{\prime(1)}$,
$\varepsilon_{P}^{(2)}$, and $\varepsilon_{P}^{\prime(2)}$, and fits into the
aforementioned rules (\ref{rule1})(\ref{rule2})(\ref{rule3}) (\ref{rule4}). It
is also worth noting that the appearance of the last term in the second
exponent of Eq.(\ref{ex2}) originates from the contraction between $\partial
X\left(  z_{2}\right) $ and $\bar{\partial} \tilde{X}\left( \bar{z}_{2}
\right) $ in Eq.(\ref{def-amp}), which is a characteristic of string D-brane scattering.


The explicit form of the amplitude for the current example is
\begin{align}
A  &  =\int_{0}^{1} d\left(  r^{2}\right)  \left(  1-r^{2}\right)
^{a_{0}^{\prime}}r^{b_{0}-1} \left(  -\dfrac{i\sqrt{\tilde{b}_{0}}}%
{-r}\right)  \left(  -\dfrac{i\sqrt{\tilde{b}_{0}}}{-r}\right)  \left(
\dfrac{i\frac{b_{0}-1}{2M_{2}}}{\left( -r\right) ^{2}} \right)  \left(
\dfrac{i\frac{b_{0}-1}{2M_{2}}}{\left(  -r \right) ^{2}} \right) \nonumber\\
&  \cdot\left[  \left(  \dfrac{i\frac{b_{0}-1}{2M_{2}}}{-r} +\dfrac
{i\frac{a_{0}}{M_{2}}}{ \left(  1-r^{2}\right)  /r } \right)  \left(
\dfrac{i\frac{b_{0}-1}{2M_{2}}}{-r} +\dfrac{i\frac{a_{0}}{M_{2}}}{ \left(
1-r^{2}\right) /r} \right)  +\dfrac{\frac{a_{0}}{M_{2}^{2}}}{\left(
1-r^{2}\right) ^{2}} \right] \label{ex-reduction1}\\
&  = -\left(  \sqrt{\tilde{b}_{0}}\right) ^{2} \left(  \frac{b_{0} -1}{2M_{2}%
}\right) ^{4} \int_{0}^{1} d\left(  r^{2}\right)  \left(  1-r^{2}\right)
^{a_{0}^{\prime}}r^{b_{0}-9}\nonumber\\
&  \cdot\left[  \left(  \sum_{l=0}^{2} \binom{2}{l}\left(  \dfrac{-r^{2}}{
\left(  1-r^{2}\right)  }\frac{2 a_{0}}{b_{0} -1}\right) ^{l} \right)
-\dfrac{r^{2}}{\left(  1-r^{2}\right) ^{2}}\frac{ 4 a_{0}}{\left(  b_{0} -1
\right) ^{2}} \right] \label{ex-reduction2}\\
&  \sim-\left(  \sqrt{\tilde{b}_{0}}\right) ^{2} \left(  \frac{b_{0}
-1}{2M_{2}}\right) ^{4} B\left(  a_{0} +1,\frac{b_{0} -7}{2}\right)
\nonumber\\
&  \cdot\left[  \left(  \sum_{l=0}^{2} \binom{2}{l}\left(  -\frac{2 }{b_{0}
-1}\right) ^{l} \left(  \frac{b_{0} -7}{2}\right) _{l} \right)  -\frac{ 4
}{\left(  b_{0} -1 \right) ^{2}} \left(  \frac{b_{0} -7}{2}\right)  \right]
\label{ex-reduction3}\\
&  = -\left(  \sqrt{\tilde{b}_{0}}\right) ^{2} \left(  \frac{b_{0} -1}{2M_{2}%
}\right) ^{4} B\left(  a_{0} +1,\frac{b_{0} -7}{2}\right) \nonumber\\
& \cdot\left[  \ _{2} F_{0} \left(  -2, \frac{b_{0} -7}{2},\frac{2 }{b_{0}
-1}\right)  -\frac{ 4 }{\left(  b_{0} -1 \right) ^{2}} \left(  \frac{b_{0}
-7}{2}\right)  \right] \label{ex-reduction4}%
\end{align}
where we have used Eq.(\ref{Regge behavior}).

\subsection{General cases}

Now we move on to general cases. The vertex operator corresponding to a
general massive state with $d$ left-modes  and $d^{\prime}$ right-modes is of
the following form.
\begin{align}
V=i^{d+d^{\prime}}\varepsilon_{\mu_{1}\cdots\mu_{d+d^{\prime}}}:\partial
^{n_{1}} X^{\mu_{1}}\cdots\partial^{n_{d}} X^{\mu_{d}}e^{ik_{2} X}\left(
z\right) :\ :\bar{\partial}^{n_{d+1}} \tilde{X}^{\mu_{d+1}}\cdots\bar
{\partial}^{n_{d+d^{\prime}}} \tilde{X}^{\mu_{d+d^{\prime}}}e^{ik_{2}
\tilde{X}}\left(  \bar{z}\right) :\label{op}%
\end{align}
The vertex operators corresponding to the states Eq.(\ref{general states}) are
expressed in this covariant form by
\begin{align*}
& d =\sum_{n>0} p_{n}+q_{n},\quad d^{\prime}=\sum_{n>0} p^{\prime}_{n} +
q^{\prime}_{n}\\
& \left(  n_{1},n_{2},\cdots,n_{d+d^{\prime}} \right)  =\left( \cdots
,\underbrace{m,\cdots,m}_{p_{m}},\cdots,\underbrace{n,\cdots,n}_{q_{n}}%
,\cdots,\underbrace{m^{\prime},\cdots,m^{\prime}}_{p^{\prime}_{m^{\prime}}%
},\cdots,\underbrace{n^{\prime},\cdots,n^{\prime}}_{q^{\prime}_{n^{\prime}}%
},\cdots\right) \\
&  \varepsilon_{\cdots\underbrace{T\cdots T}_{p_{m}}\cdots\underbrace{P\cdots
P}_{q_{n}}\cdots\underbrace{T\cdots T}_{p^{\prime}_{m^{\prime}}}%
\cdots\underbrace{P\cdots P}_{q^{\prime}_{n^{\prime}}}\cdots}=1.
\end{align*}
For the calculation of the correlator involving the operator Eq.(\ref{op}), we
introduce parameters associated with the polarization tensor and exponentiate
the kinematic factors.
\begin{align*}
\varepsilon_{TTT\cdots PPP\cdots TTT\cdots PPP\cdots}  & \rightarrow
\prod_{n>0} \prod_{i=1}^{p_{n}} \prod_{j=1}^{q_{n}}\prod_{i^{\prime}%
=1}^{p^{\prime}_{n}}\prod_{j^{\prime}=1}^{q^{\prime}_{n}} \varepsilon_{T_{i}%
}^{(n)} \varepsilon_{P_{j}}^{(n)} \varepsilon_{T_{i^{\prime}}}^{\prime(n)}
\varepsilon_{P_{j^{\prime}}}^{\prime(n)}%
\end{align*}
\begin{align}
V =  &  \left( i\right) ^{\sum_{n>0}p_{n} + p^{\prime}_{n} + q_{n} +q^{\prime
}_{n}} \left[  :\exp\left\{  ik_{2} X(z) + \sum_{n>0} \sum_{i=1}^{p_{n}}
\varepsilon_{T_{i}}^{(n)}\partial^{n} X^{T} (z) + \sum_{m>0} \sum_{j=1}%
^{q_{m}} \varepsilon_{P_{j}}^{(m)}\partial^{m} X^{P} (z) \right\} : \right.
\nonumber\\
&  \quad\quad\times\left.  :\exp\left\{  ik_{2} \tilde{X}(\bar{z}) +
\sum_{n>0} \sum_{i=1}^{p^{\prime}_{n}} \varepsilon_{T_{i}}^{\prime(n)}%
\partial^{n} \tilde{X}^{T} (\bar{z}) + \sum_{m>0} \sum_{j=1}^{q^{\prime}_{m}}
\varepsilon_{P_{j}}^{\prime(m)}\partial^{m} \tilde{X}^{P} (\bar{z}) \right\} :
\right] _{\mathrm{linear\ terms}}\label{exponentiation}%
\end{align}
where ``linear terms'' means the terms linear in all of $\varepsilon_{T_{i}%
}^{(n)}, \varepsilon_{P_{j}}^{(m)}, \varepsilon_{T_{i}}^{\prime(n)}$, and
$\varepsilon_{P_{j}}^{\prime(m)}$. Below we use symbols like
\begin{align*}
\varepsilon_{T^{3} P^{2} T P^{3}} \equiv\varepsilon_{T_{1}}^{(1)}%
\varepsilon_{T_{1}}^{(3)}\varepsilon_{T_{2}}^{(3)} \varepsilon_{P_{1}}%
^{(2)}\varepsilon_{P_{1}}^{(5)}\varepsilon_{T_{1}}^{\prime(1)} \varepsilon
_{P_{1}}^{\prime(1)}\varepsilon_{P_{2}}^{\prime(1)}\varepsilon_{P_{1}}%
^{\prime(2)}, \qquad\varepsilon_{T}\sum_{n} p_{n}\equiv\sum_{n>0} \sum
_{i=1}^{p_{n}}\varepsilon_{T_{i}}^{(n)}%
\end{align*}
(the meanings of these symbols are not unique.) and do not write the normal
ordering symbol : : to avoid messy expressions.

The string D-particle scattering amplitudes of these string states can be
calculated to be%
\begin{align}
A  &  =\int d^{2}z_{1}d^{2}z_{2}\cdot\varepsilon_{T^{\sum p_{n}}P^{\sum q_{n}%
}T^{\sum p^{\prime}_{n}}P^{\sum q^{\prime}_{n}}}\\
& \quad\cdot\left\langle
\begin{array}
[c]{c}%
e^{ik_{1}X}\left(  z_{1}\right)  e^{ik_{1}\tilde{X}}\left(  \bar{z}%
_{1}\right)  \cdot\prod\limits_{n>0}\left(  i\partial^{n}X^{T}\right)
^{p_{n}}\prod\limits_{m>0}\left(  i\partial^{m}X^{P}\right)  ^{q_{m}}%
e^{ik_{2}X}\left(  z_{2}\right) \\
\cdot\prod\limits_{n>0}\left(  i\bar{\partial}^{n}\tilde{X}^{T}\right)
^{p_{n}^{\prime}}\prod\limits_{m>0}\left(  i\bar{\partial}^{m}\tilde{X}%
^{P}\right)  ^{q_{m}^{\prime}}e^{ik_{2}\tilde{X}}\left(  \bar{z}_{2}\right)
\end{array}
\right\rangle \nonumber\\
&  \equiv(i)^{\sum\limits_{n>0}p_{n}+p_{n}^{\prime}+q_{n}+q_{n}^{\prime}%
}A^{\prime}\label{phase}\\
&  =(i)^{\sum\limits_{n>0}p_{n}+p_{n}^{\prime}+q_{n}+q_{n}^{\prime}}\int
d^{2}z_{1}d^{2}z_{2}\nonumber\\
&  \cdot\exp\left\{
\begin{array}
[c]{c}%
\left\langle \left(  ik_{1}X\right)  \left(  z_{1}\right)  \left(
ik_{1}\tilde{X}\right)  \left(  \bar{z}_{1}\right)  \right\rangle \\
+\left\langle
\begin{array}
[c]{c}%
\left(  \varepsilon_{T}\sum\limits_{n>0}p_{n}\partial^{n}X^{T}+\varepsilon
_{P}\sum\limits_{m>0}q_{m}\partial^{m}X^{P}+ik_{2}X\right)  \left(
z_{2}\right) \\
\left(  \varepsilon_{T}^{\prime}\sum\limits_{n>0}p_{n}^{\prime}\bar{\partial
}^{n}\tilde{X}^{T}+\varepsilon_{P}^{\prime}\sum\limits_{m>0}q_{m}^{\prime}%
\bar{\partial}^{m}\tilde{X}^{P}+ik_{2}\tilde{X}\right)  \left(  \bar{z}%
_{2}\right)
\end{array}
\right\rangle \\
+\left\langle \left(  ik_{1}X\right)  \left(  z_{1}\right)  \left(
\varepsilon_{T}\sum\limits_{n>0}p_{n}\partial^{n}X^{T}+\varepsilon_{P}%
\sum\limits_{m>0}q_{m}\partial^{m}X^{P}+ik_{2}X\right)  \left(  z_{2}\right)
\right\rangle \\
+\left\langle \left(  ik_{1}\tilde{X}\right)  \left(  \bar{z}_{1}\right)
\left(  \varepsilon_{T}^{\prime}\sum\limits_{n>0}p_{n}^{\prime}\bar{\partial
}^{n}\tilde{X}^{T}+\varepsilon_{P}^{\prime}\sum\limits_{m>0}q_{m}^{\prime}%
\bar{\partial}^{m}\tilde{X}^{P}+ik_{2}\tilde{X}\right)  \left(  \bar{z}%
_{2}\right)  \right\rangle \\
+\left\langle \left(  ik_{1}X\right)  \left(  z_{1}\right)  \left(
\varepsilon_{T}^{\prime}\sum\limits_{n>0}p_{n}^{\prime}\bar{\partial}%
^{n}\tilde{X}^{T}+\varepsilon_{P}^{\prime}\sum\limits_{m>0}q_{m}^{\prime}%
\bar{\partial}^{m}\tilde{X}^{P}+ik_{2}\tilde{X}\right)  \left(  \bar{z}%
_{2}\right)  \right\rangle \\
+\left\langle \left(  ik_{1}\tilde{X}\right)  \left(  \bar{z}_{1}\right)
\left(  \varepsilon_{T}\sum\limits_{n>0}p_{n}\partial^{n}X^{T}+\varepsilon
_{P}\sum\limits_{m>0}q_{m}\partial^{m}X^{P}+ik_{2}X\right)  \left(
z_{2}\right)  \right\rangle
\end{array}
\right\}  \label{GG}%
\end{align}
where only linear terms are taken in the expansion of the exponential (in the
sense of Eq.(\ref{exponentiation})). In Eq.(\ref{GG}), we have used the
simplified notation $\varepsilon_{T_{j}}^{(n)}\equiv\varepsilon_{T},$
$j=1,2,...p_{n}$, $n\in Z_{+}$ for the spin polarizations, and similarly for
the other polarizations. The exact meanings of the summations in the exponent
are the ones like Eq.(\ref{exponentiation}). Note that there will be terms
corresponding to quadratic in the spin polarization. The amplitude $A^{\prime
}$ can be reduced to
\begin{align}
A^{\prime}  &  =\int d^{2}z_{1}d^{2}z_{2}\left\langle e^{ik_{1}X}\left(
z_{1}\right)  e^{ik_{1}\tilde{X}}\left(  \bar{z}_{1}\right)  e^{ik_{2}%
X}\left(  z_{2}\right)  e^{ik_{2}\tilde{X}}\left(  \bar{z}_{2}\right)
\right\rangle \nonumber\\
&  \cdot\exp\left\{
\begin{array}
[c]{l}%
-\varepsilon_{T}\sum\limits_{n>0}p_{n}\left[
\begin{array}
[c]{c}%
ie^{T}\cdot k_{1}\partial_{2}^{n}\left\langle X\left(  z_{1}\right)  X\left(
z_{2}\right)  \right\rangle +ie^{T}\cdot D\cdot k_{1}\partial_{2}%
^{n}\left\langle \tilde{X}\left(  \bar{z}_{1}\right)  X\left(  z_{2}\right)
\right\rangle \\
+ie^{T}\cdot D\cdot k_{2}\partial_{2}^{n}\left\langle \tilde{X}\left(  \bar
{z}_{2}\right)  X\left(  z_{2}\right)  \right\rangle
\end{array}
\right] \\
-\varepsilon_{T}^{\prime}\sum\limits_{n^{\prime}>0}p_{n^{\prime}}^{\prime
}\left[
\begin{array}
[c]{c}%
ie^{T}\cdot D\cdot k_{1}\bar{\partial}_{2}^{n^{\prime}}\left\langle X\left(
z_{1}\right)  \tilde{X}\left(  \bar{z}_{2}\right)  \right\rangle +ie^{T}\cdot
k_{1}\bar{\partial}_{2}^{n^{\prime}}\left\langle \tilde{X}\left(  \bar{z}%
_{1}\right)  \tilde{X}\left(  \bar{z}_{2}\right)  \right\rangle \\
+ie^{T}\cdot D\cdot k_{2}\bar{\partial}_{2}^{n^{\prime}}\left\langle X\left(
z_{2}\right)  \tilde{X}\left(  \bar{z}_{2}\right)  \right\rangle
\end{array}
\right] \\
-\varepsilon_{P}\sum\limits_{m>0}q_{m}\left[
\begin{array}
[c]{c}%
ie^{P}\cdot k_{1}\partial_{2}^{m}\left\langle X\left(  z_{1}\right)  X\left(
z_{2}\right)  \right\rangle +ie^{P}\cdot D\cdot k_{1}\partial_{2}%
^{m}\left\langle \tilde{X}\left(  \bar{z}_{1}\right)  X\left(  z_{2}\right)
\right\rangle \\
+ie^{P}\cdot D\cdot k_{2}\partial_{2}^{m}\left\langle \tilde{X}\left(  \bar
{z}_{2}\right)  X\left(  z_{2}\right)  \right\rangle
\end{array}
\right] \\
-\varepsilon_{P}^{\prime}\sum\limits_{m^{\prime}>0}q_{m^{\prime}}^{\prime
}\left[
\begin{array}
[c]{c}%
ie^{P}\cdot D\cdot k_{1}\bar{\partial}_{2}^{m^{\prime}}\left\langle X\left(
z_{1}\right)  \tilde{X}\left(  \bar{z}_{2}\right)  \right\rangle +ie^{P}\cdot
k_{1}\bar{\partial}_{2}^{m^{\prime}}\left\langle \tilde{X}\left(  \bar{z}%
_{1}\right)  \tilde{X}\left(  \bar{z}_{2}\right)  \right\rangle \\
+ie^{P}\cdot D\cdot k_{2}\bar{\partial}_{2}^{m^{\prime}}\left\langle X\left(
z_{2}\right)  \tilde{X}\left(  \bar{z}_{2}\right)  \right\rangle
\end{array}
\right] \\
-\varepsilon_{T}\varepsilon_{T}^{\prime}\sum\limits_{n,n^{\prime}>0}%
p_{n}p_{n^{\prime}}^{\prime}\left(  e^{T}\cdot D\cdot e^{T}\right)
\partial^{n}\bar{\partial}^{n^{\prime}}\left\langle X\left(  z_{2}\right)
\tilde{X}\left(  \bar{z}_{2}\right)  \right\rangle \\
-\varepsilon_{P}\varepsilon_{P}^{\prime}\sum\limits_{m,m^{\prime}>0}%
q_{m}q_{m^{\prime}}^{\prime}\left(  e^{P}\cdot D\cdot e^{P}\right)
\partial^{m}\bar{\partial}^{m^{\prime}}\left\langle X\left(  z_{2}\right)
\tilde{X}\left(  \bar{z}_{2}\right)  \right\rangle \\
-\varepsilon_{T}\varepsilon_{P}^{\prime}\sum\limits_{n,m^{\prime}>0}%
p_{n}q_{m^{\prime}}^{\prime}\left(  e^{T}\cdot D\cdot e^{P}\right)
\partial^{n}\bar{\partial}^{m^{\prime}}\left\langle X\left(  z_{2}\right)
\tilde{X}\left(  \bar{z}_{2}\right)  \right\rangle \\
-\varepsilon_{P}\varepsilon_{T}^{\prime}\sum\limits_{n^{\prime},m>0}%
q_{m}p_{n^{\prime}}^{\prime}\left(  e^{P}\cdot D\cdot e^{T}\right)
\partial^{m}\bar{\partial}^{n^{\prime}}\left\langle X\left(  z_{2}\right)
\tilde{X}\left(  \bar{z}_{2}\right)  \right\rangle
\end{array}
\right\}
\end{align}
where only linear terms are taken in the expansion of the exponential. We can
now put in the propagators in Eq.(\ref{DD1}) to Eq.(\ref{DD}) to get%
\begin{align}
&  A^{\prime}=\int d^{2}z_{1}d^{2}z_{2}\left(  1-z_{1}\bar{z}_{1}\right)
^{a_{0}}\left(  1-z_{2}\bar{z}_{2}\right)  ^{a_{0}^{\prime}}\left\vert
z_{1}-z_{2}\right\vert ^{b_{0}-1}\left\vert 1-z_{1}\bar{z}_{2}\right\vert
^{c_{0}-1}\nonumber\\
&  \exp\left\{
\begin{array}
[c]{l}%
\varepsilon_{T}\sum\limits_{n>0}p_{n}\left[  \dfrac{i\left(  n-1\right)
!e^{T}\cdot k_{1}}{\left(  z_{1}-z_{2}\right)  ^{n}}+\dfrac{i\left(
n-1\right)  !e^{T}\cdot D\cdot k_{1}}{\left(  1-\bar{z}_{1}z_{2}\right)  ^{n}%
}\bar{z}_{1}^{n}+\dfrac{i\left(  n-1\right)  !e^{T}\cdot D\cdot k_{2}}{\left(
1-\bar{z}_{2}z_{2}\right)  ^{n}}\bar{z}_{2}^{n}\right] \\
+\varepsilon_{T}^{\prime}\sum\limits_{n^{\prime}>0}p_{n^{\prime}}^{\prime
}\left[  \dfrac{i\left(  n^{\prime}-1\right)  !e^{T}\cdot D\cdot k_{1}%
}{\left(  1-z_{1}\bar{z}_{2}\right)  ^{n^{\prime}}}z_{1}^{n^{\prime}}%
+\dfrac{i\left(  n^{\prime}-1\right)  !e^{T}\cdot k_{1}}{\left(  \bar{z}%
_{1}-\bar{z}_{2}\right)  ^{n^{\prime}}}+\dfrac{i\left(  n^{\prime}-1\right)
!e^{T}\cdot D\cdot k_{2}}{\left(  1-z_{2}\bar{z}_{2}\right)  ^{n^{\prime}}%
}z_{2}^{n^{\prime}}\right] \\
+\varepsilon_{P}\sum\limits_{m>0}q_{m}\left[  \dfrac{i\left(  m-1\right)
!e^{P}\cdot k_{1}}{\left(  z_{1}-z_{2}\right)  ^{m}}+\dfrac{i\left(
m-1\right)  !e^{P}\cdot D\cdot k_{1}}{\left(  1-\bar{z}_{1}z_{2}\right)  ^{m}%
}\bar{z}_{1}^{m}+\dfrac{i\left(  m-1\right)  !e^{P}\cdot D\cdot k_{2}}{\left(
1-\bar{z}_{2}z_{2}\right)  ^{m}}\bar{z}_{2}^{m}\right] \\
+\varepsilon_{P}^{\prime}\sum\limits_{m^{\prime}>0}q_{m^{\prime}}^{\prime
}\left[
\begin{array}
[c]{c}%
\dfrac{i\left(  m^{\prime}-1\right)  !e^{P}\cdot D\cdot k_{1}}{\left(
1-z_{1}\bar{z}_{2}\right)  ^{m^{\prime}}}z_{1}^{m^{\prime}}+\dfrac{i\left(
m^{\prime}-1\right)  !e^{P}\cdot k_{1}}{\left(  \bar{z}_{1}-\bar{z}%
_{2}\right)  ^{m^{\prime}}}\\
+\dfrac{i\left(  m^{\prime}-1\right)  !e^{P}\cdot D\cdot k_{2}}{\left(
1-z_{2}\bar{z}_{2}\right)  ^{m^{\prime}}}z_{2}^{m^{\prime}}%
\end{array}
\right] \\
-\varepsilon_{T}\varepsilon_{T}^{\prime}\sum\limits_{n,n^{\prime}>0}%
p_{n}p_{n^{\prime}}^{\prime}\left(  e^{T}\cdot D\cdot e^{T}\right)
\partial^{n}\bar{\partial}^{n^{\prime}} \ln\left(  1-z_{2} \bar{z}_{2}\right)
\\
-\varepsilon_{P}\varepsilon_{P}^{\prime}\sum\limits_{m,m^{\prime}>0}%
q_{m}q_{m^{\prime}}^{\prime}\left(  e^{P}\cdot D\cdot e^{P}\right)
\partial^{m}\bar{\partial}^{m^{\prime}} \ln\left(  1-z_{2} \bar{z}_{2}\right)
\\
-\varepsilon_{T}\varepsilon_{P}^{\prime}\sum\limits_{n,m^{\prime}>0}%
p_{n}q_{m^{\prime}}^{\prime}\left(  e^{T}\cdot D\cdot e^{P}\right)
\partial^{n}\bar{\partial}^{m^{\prime}} \ln\left(  1-z_{2} \bar{z}_{2}\right)
\\
-\varepsilon_{P}\varepsilon_{T}^{\prime}\sum\limits_{n^{\prime},m>0}%
q_{m}p_{n^{\prime}}^{\prime}\left(  e^{P}\cdot D\cdot e^{T}\right)
\partial^{m}\bar{\partial}^{n^{\prime}} \ln\left(  1-z_{2} \bar{z}_{2}\right)
\\
\end{array}
\right\} \label{eq1}%
\end{align}
where only linear terms are taken in the expansion of the exponential. To fix
the $SL(2,R)$ modulus group on the disk, we set $z_{1}=0$ and $z_{2}=r$, then
$d^{2}z_{1}d^{2}z_{2}=d\left(  r^{2}\right)  .$ By using Eq.(\ref{Kine}), the
amplitude can then be reduced to%
\begin{align}
A^{\prime}  &  =\int d\left(  r^{2}\right)  \left(  1-r^{2}\right)
^{a_{0}^{\prime}}r^{b_{0}-1}\nonumber\\
&  \exp\left\{
\begin{array}
[c]{l}%
\varepsilon_{T}\sum\limits_{n>0}p_{n}\left[  -\dfrac{i\left(  n-1\right)
!\sqrt{\tilde{b}_{0}}}{(-r)^{n}}\right]  +\varepsilon_{T}^{\prime}%
\sum\limits_{n^{\prime}>0}p_{n^{\prime}}^{\prime}\left[  -\dfrac{i\left(
n^{\prime}-1\right)  !\sqrt{\tilde{b}_{0}}}{(-r)^{n^{\prime}}}\right] \\
+\varepsilon_{P}\sum\limits_{m>0}q_{m}\left[  \dfrac{i\left(  m-1\right)
!\frac{b_{0}-1}{2M_{2}}}{(-r)^{m}}+\dfrac{i\left(  m-1\right)  !\frac{a_{0}%
}{M_{2}}}{\left[  \left(  1-r^{2}\right)  /r\right]  ^{m}}\right] \\
+\varepsilon_{P}^{\prime}\sum\limits_{m^{\prime}>0}q_{m^{\prime}}^{\prime
}\left[  \dfrac{i\left(  m^{\prime}-1\right)  !\frac{b_{0}-1}{2M_{2}}%
}{(-r)^{m^{\prime}}}+\dfrac{i\left(  m^{\prime}-1\right)  !\frac{a_{0}}{M_{2}%
}}{\left[  \left(  1-r^{2}\right)  /r\right]  ^{m^{\prime}}}\right] \\
-\varepsilon_{T}\varepsilon_{T}^{\prime}\sum\limits_{n,n^{\prime}>0}%
p_{n}p_{n^{\prime}}^{\prime}\partial^{n}\bar{\partial}^{n^{\prime}} \ln\left(
1-z_{2} \bar{z}_{2}\right) \big|_{z_{2}=\bar{z}_{2}=r}\\
-\varepsilon_{P}\varepsilon_{P}^{\prime}\sum\limits_{m,m^{\prime}>0}%
q_{m}q_{m^{\prime}}^{\prime}\partial^{m}\bar{\partial}^{m^{\prime}} \ln\left(
1-z_{2} \bar{z}_{2}\right) \big|_{z_{2}=\bar{z}_{2}=r} \frac{a_{0}}{M_{2}^{2}}%
\end{array}
\right\} \label{reduction1}%
\end{align}
where only linear terms are taken in the expansion of the exponential.

Now we use the energy counting (\ref{rule5}) and show how we reach the rules
(\ref{rule1})(\ref{rule2})(\ref{rule3})(\ref{rule4}). We can see immediately
that in the exponent of Eq.(\ref{reduction1}), the terms linear in
$\varepsilon^{ (n)}_{P_{i}}$ or $\varepsilon^{\prime(n)}_{P_{i}}$ are
dominated by their first terms if $m\ge2$ or $m^{\prime}\ge2$. We can see also
that most of the terms in the forth and fifth lines of the exponent are
discarded as subleading. If we start with the terms consisting of only the
factors coming from the first three lines, the other terms are obtained by
series of replacements of two factors in them with one factors coming from the
forth and fifth lines, and for each of the replacements we can see how it
changes the power of energy. We do not need to calculate the infinite number
of derivatives. For each differentiation the increase of the power of $1/r$ is
less than or equal to 1, while the powers of $1/r$ in the first three lines
increase with $n,n^{\prime},m$ or $m^{\prime}$, which implies that if one term
in the forth or fifth line is discarded, the terms with higher $n,n^{\prime
},m, m^{\prime}$ in the same line are also discarded. The sequences of those
discarded terms start at $\left( n,n^{\prime}\right) =\left( 1,1\right) $,
$\left( m,m^{\prime}\right)  =\left( 1,2\right) $, and $\left( m,m^{\prime
}\right) =\left( 2,1\right) $. In this way, we can see that only the terms
with $m=m^{\prime}=1$ in the fifth line contribute to the leading behavior.
Thus we obtain the generalization of Eq.(\ref{ex2})
\begin{align}
&  A^{\prime}=\int d\left(  r^{2}\right)  \left(  1-r^{2}\right)
^{a_{0}^{\prime}}r^{b_{0}-1}\nonumber\\
&  \exp\left\{
\begin{array}
[c]{l}%
\varepsilon_{T}\sum\limits_{n>0}p_{n}\left[  -\dfrac{i\left(  n-1\right)
!\sqrt{\tilde{b}_{0}}}{(-r)^{n}}\right]  +\varepsilon_{T}^{\prime}%
\sum\limits_{n^{\prime}>0}p_{n^{\prime}}^{\prime}\left[  -\dfrac{i\left(
n^{\prime}-1\right)  !\sqrt{\tilde{b}_{0}}}{(-r)^{n^{\prime}}}\right] \\
+\varepsilon_{P}\sum\limits_{m>1}q_{m}\left[  \dfrac{i\left(  m-1\right)
!\frac{b_{0}-1}{2M_{2}}}{(-r)^{m}}\right]  +\varepsilon_{P}^{\prime}%
\sum\limits_{m^{\prime}>1}q_{m^{\prime}}^{\prime}\left[  \dfrac{i\left(
m^{\prime}-1\right)  !\frac{b_{0}-1}{2M_{2}}}{(-r)^{m^{\prime}}}\right]
\end{array}
\right\}  _{ \varepsilon_{T^{\sum p_{n}} P^{\sum^{\prime}q_{n}}T^{\sum
p^{\prime}_{n}}P^{\sum^{\prime}q^{\prime}_{n}}}}\nonumber\\
&  \exp\left\{  \varepsilon_{P}q_{1}\left[  \dfrac{i\frac{b_{0}-1}{2M_{2}}%
}{-r}+\dfrac{i\frac{a_{0}}{M_{2}}r}{1-r^{2}}\right]  +\varepsilon_{P}^{\prime
}q_{1}^{\prime}\left[  \dfrac{i\frac{b_{0}-1}{2M_{2}}}{-r}+\dfrac{i\frac
{a_{0}}{M_{2}}r}{1-r^{2}}\right]  +\varepsilon_{P}\varepsilon_{P}^{\prime
}q_{1}q_{1}^{\prime}\dfrac{\frac{a_{0}}{M_{2}^{2}}}{\left(  1-r^{2}\right)
^{2}}\right\}  _{\varepsilon_{P^{q_{1}}P^{q_{1}^{\prime}}}} \label{drop}%
\end{align}
where the symbols $\varepsilon_{\cdots}$ are similar to the ones in
Eq.(\ref{ex2}) and indicate that we take the coefficients of the products of
the dummy variables in the exponents. ( $\varepsilon^{(1)}_{P_{i}}$ and
$\varepsilon^{\prime(1)}_{P_{i}}$ are excluded in the ``sums'' $\sum^{\prime}%
$.) Note that the last term in the last line of Eq.(\ref{drop}) is quadratic
in the polarization. This term is a characteristic of string D-brane
scattering and has no analog in any of the previous works. It will play a
crucial role in the following calculation in this paper.

For further calculation, we first note that%
\begin{align}
&  \exp\left\{  \varepsilon_{P}q_{1}\left[  \dfrac{i\frac{b_{0}-1}{2M_{2}}%
}{-r}+\dfrac{i\frac{a_{0}}{M_{2}}r}{1-r^{2}}\right]  +\varepsilon_{P}^{\prime
}q_{1}^{\prime}\left[  \dfrac{i\frac{b_{0}-1}{2M_{2}}}{-r}+\dfrac{i\frac
{a_{0}}{M_{2}}r}{1-r^{2}}\right]  +\varepsilon_{P}\varepsilon_{P}^{\prime
}q_{1}q_{1}^{\prime}\dfrac{\frac{a_{0}}{M_{2}^{2}}}{\left(  1-r^{2}\right)
^{2}}\right\}  _{\varepsilon_{P^{q_{1}}P^{q_{1}^{\prime}}}}\nonumber\\
&  =\varepsilon_{P^{q_{1}}P^{q_{1}^{\prime}}}\sum_{j=0}^{\min\left\{
q_{1},q_{1}^{\prime}\right\}  }\binom{q_{1}}{j}\binom{q_{1}^{\prime}}%
{j}j!\left(  \dfrac{i\frac{b_{0}-1}{2M_{2}}}{-r}+\dfrac{i\frac{a_{0}}{M_{2}}%
r}{1-r^{2}}\right)  ^{q_{1}+q_{1}^{\prime}-2j}\left(  \dfrac{\frac{a_{0}%
}{M_{2}^{2}}}{\left(  1-r^{2}\right)  ^{2}}\right)  ^{j}.
\end{align}
Thus the amplitude can be further reduced to%
\begin{align}
A^{\prime}  &  =\int d\left(  r^{2}\right)  \left(  1-r^{2}\right)
^{a_{0}^{\prime}}r^{b_{0}-1}\nonumber\\
&  \cdot\prod\limits_{n>0}\left[  -\dfrac{i\left(  n-1\right)  !\sqrt
{\tilde{b}_{0}}}{(-r)^{n}}\right]  ^{p_{n}}\prod\limits_{n^{\prime}>0}\left[
-\dfrac{i\left(  n^{\prime}-1\right)  !\sqrt{\tilde{b}_{0}}}{(-r)^{n^{\prime}%
}}\right]  ^{p_{n^{\prime}}^{\prime}}\nonumber\\
&  \cdot\prod\limits_{m>1}\left[  \dfrac{i\left(  m-1\right)  !\frac{b_{0}%
-1}{2M_{2}}}{(-r)^{m}}\right]  ^{q_{m}}\prod\limits_{m^{\prime}>1}\left[
\dfrac{i\left(  m^{\prime}-1\right)  !\frac{b_{0}-1}{2M_{2}}}{(-r)^{m^{\prime
}}}\right]  ^{q_{m^{\prime}}}\nonumber\\
&  \cdot\sum_{j=0}^{\min\left\{  q_{1},q_{1}^{\prime}\right\}  }\sum
_{l=0}^{q_{1}+q_{1}^{\prime}-2j}j!\binom{q_{1}}{j}\binom{q_{1}^{\prime}}%
{j}\binom{q_{1}+q_{1}^{\prime}-2j}{l}\nonumber\\
&  \cdot\left(  \dfrac{i\frac{b_{0}-1}{2M_{2}}}{-r}\right)  ^{q_{1}%
+q_{1}^{\prime}-2j-l}\left(  \dfrac{i\frac{a_{0}}{M_{2}}r}{1-r^{2}}\right)
^{l}\left(  \dfrac{\frac{a_{0}}{M_{2}^{2}}}{\left(  1-r^{2}\right)  ^{2}%
}\right)  ^{j},
\end{align}
which, in the case of the state (\ref{example-state}), is reduced to
Eq.(\ref{ex-reduction2}). We can now do the integration to get%
\begin{align}
A^{\prime}  &  =\left(  i\frac{b_{0}-1}{2M_{2}}\right)  ^{q_{1}+q_{1}^{\prime
}}\cdot\prod\limits_{n>0}\left(  \left[  -i\left(  n-1\right)  !\sqrt
{\tilde{b}_{0}}\right]  ^{p_{n}}\left[  -i\left(  n-1\right)  !\sqrt{\tilde
{b}_{0}}\right]  ^{p_{n}^{\prime}}\right) \nonumber\\
&  \cdot\prod\limits_{m>1}\left(  \left[  i\left(  m-1\right)  !\frac{b_{0}%
-1}{2M_{2}}\right]  ^{q_{m}}\left[  i\left(  m-1\right)  !\frac{b_{0}%
-1}{2M_{2}}\right]  ^{q_{m}}\right) \nonumber\\
&  \cdot\sum_{j=0}^{\min\left\{  q_{1},q_{1}^{\prime}\right\}  }\sum
_{l=0}^{q_{1}+q_{1}^{\prime}-2j}j!\binom{q_{1}}{j}\binom{q_{1}^{\prime}}%
{j}\binom{q_{1}+q_{1}^{\prime}-2j}{l}\left(  \frac{-2}{b_{0}-1}\right)
^{l}\left(  \frac{-4}{\left(  b_{0}-1\right)  ^{2}}\right)  ^{j}\nonumber\\
&  \qquad\cdot B\left(  a_{0}+1,\frac{b_{0}+1-N}{2}\right)  \left(
\frac{b_{0}+1-N}{2}\right)  _{j}\left(  \frac{b_{0}+1-N}{2}+j\right)  _{l}
\label{min}%
\end{align}
where we have done the expansion of the beta function in the RR as following
\begin{align}
&  B\left(  a^{\prime}_{0}+1-l-2j,\dfrac{b_{0}+1-N}{2}+l+j\right) \nonumber\\
&  \approx B\left(  a_{0}+1,\frac{b_{0}+1-N}{2}\right)  \frac{\left(
\frac{b_{0}+1-N}{2}\right)  _{l+j}}{a_{0}^{l+j}}\nonumber\\
&  =B\left(  a_{0}+1,\frac{b_{0}+1-N}{2}\right)  \frac{\left(  \frac
{b_{0}+1-N}{2}\right)  _{j}\left(  \frac{b_{0}+1-N}{2}+j\right)  _{l}}%
{a_{0}^{l+j}}.
\end{align}
Note that in the case of the state (\ref{example-state}), Eq.(\ref{min}) is
reduced to Eq.(\ref{ex-reduction3}). Performing the summation over $n$, we
obtain%
\begin{align}
A^{\prime}  &  =\left(  i\frac{b_{0}-1}{2M_{2}}\right)  ^{q_{1}+q_{1}}%
\cdot\prod\limits_{n>0}\left(  \left[  -i\left(  n-1\right)  !\sqrt{\tilde
{b}_{0}}\right]  ^{p_{n}+p_{n}^{\prime}}\right)  \prod\limits_{m>1}\left(
\left[  i\left(  m-1\right)  !\frac{b_{0}-1}{2M_{2}}\right]  ^{q_{m}%
+q_{m}^{\prime}}\right) \nonumber\\
&  \cdot B\left(  a_{0}+1,\frac{b_{0}+1-N}{2}\right)  \sum_{j=0}^{\min\left\{
q_{1},q_{1}^{\prime}\right\}  }(-1)^{j}j!\binom{q_{1}}{j}\binom{q_{1}^{\prime
}}{j}\left(  \frac{b_{0}+1-N}{2}\right)  _{j}\left(  \frac{2}{b_{0}-1}\right)
^{2j}\nonumber\\
&  \cdot_{2}F_{0}\left(  -q_{1}-q_{1}^{\prime}+2j,\frac{b_{0}+1-N}{2}%
+j,\frac{2}{b_{0}-1}\right) ,
\end{align}
which, in the case of the state (\ref{example-state}), is reduced to
Eq.(\ref{ex-reduction4}). Finally we can use the identity of the Kummer
function%
\begin{align}
&  2^{2m}\ \tilde{t}^{-2m}U\left(  -2m,\frac{t}{2}+2-2m,\frac{\tilde{t}}%
{2}\right) \nonumber\\
&  =\,_{2}F_{0}\left(  -2m,-1-\frac{t}{2},-\frac{2}{\tilde{t}}\right)
\nonumber\\
&  \equiv\sum_{j=0}^{2m}\left(  -2m\right)  _{j}\left(  -1-\frac{t}{2}\right)
_{j}\frac{\left(  -\frac{2}{\tilde{t}}\right)  ^{j}}{j!}\nonumber\\
&  =\sum_{j=0}^{2m}{\binom{2m}{j}}\left(  -1-\frac{t}{2}\right)  _{j}\left(
\frac{2}{\tilde{t}}\right)  ^{j} \label{Kummer}%
\end{align}
to get the final form of the amplitude%
\begin{align}
A^{\prime}  &  =\prod\limits_{n>0}\left(  \left[  -i\left(  n-1\right)
!\sqrt{\tilde{b}_{0}}\right]  ^{p_{n}+p_{n}^{\prime}}\right)  \prod
\limits_{m>1}\left(  \left[  i\left(  m-1\right)  !\frac{b_{0}-1}{2M_{2}%
}\right]  ^{q_{m}+q_{m}^{\prime}}\right)  \left(  -\frac{i}{M_{2}}\right)
^{q_{1}+q_{1}^{\prime}}\nonumber\\
&  \cdot B\left(  a_{0}+1,\frac{b_{0}+1-N}{2}\right)  \sum_{j=0}^{\min\left\{
q_{1},q_{1}^{\prime}\right\}  }(-1)^{j}j!\binom{q_{1}}{j}\binom{q_{1}^{\prime
}}{j}\left(  \frac{b_{0}+1-N}{2}\right)  _{j}\nonumber\\
&  \cdot U\left(  -q_{1}-q_{1}^{\prime}+2j,\frac{-b_{0}+N+1}{2}-q_{1}%
-q_{1}^{\prime}+j,-\frac{b_{0}-1}{2}\right)  . \label{RRamp}%
\end{align}
Note that the amplitudes in Eq.(\ref{RRamp}) can not be factorized into two
open string D-particle scattering amplitudes as in the case of closed
string-string scattering amplitudes \cite{Closed,bosonic2}. In Eq.(\ref{RRamp}%
) $U$ is the Kummer function of the second kind and is defined to be%
\begin{equation}
U(a,c,x)=\frac{\pi}{\sin\pi c}\left[  \frac{M(a,c,x)}{(a-c)!(c-1)!}%
-\frac{x^{1-c}M(a+1-c,2-c,x)}{(a-1)!(1-c)!}\right]  \text{ \ }(c\neq2,3,4...)
\end{equation}
where $M(a,c,x)=\sum_{j=0}^{\infty}\frac{(a)_{j}}{(c)_{j}}\frac{x^{j}}{j!}$ is
the Kummer function of the first kind. Note that the second argument of Kummer
function $c=c(b_{0}),$ and is not a constant as in the usual case. As a
result, $U$ as a function of $b_{0}$ is not a solution of the Kummer equation.

An interesting application of Eq.(\ref{RRamp}) is the universal power law
behavior of the amplitudes. We first define the Mandelstam variables as
$s=2E^{2}$ and $t=-(k_{1}+k_{2})^{2}.$ The second argument of the beta
function in Eq.(\ref{RRamp}) can be calculated to be%
\begin{equation}
\frac{b_{0}+1-N}{2}=\frac{2k_{1}\cdot k_{2}+1+1-N}{2}=\frac{(k_{1}+k_{2}%
)^{2}-k_{1}^{2}-k_{2}^{2}+2-N}{2}=\frac{-t-2}{2}%
\end{equation}
where we have used Eq.(\ref{II4}) and $M_{2}^{2}=(N-2).$ The amplitudes thus
give the universal power-law behavior for string states at \textit{all} mass
levels%
\begin{equation}
A\sim s^{\alpha(t)}\text{ \ (in the RR)}%
\end{equation}
where
\begin{equation}
\alpha(t)=a(0)+\alpha^{\prime}t\text{, \ }a(0)=1\text{ and }\alpha^{\prime
}=\frac{1}{2}.
\end{equation}
%

\setcounter{equation}{0}
\renewcommand{\theequation}{\arabic{section}.\arabic{equation}}%

\section{Ratios on the Fixed Angle Regime}

\bigskip We begin with a brief review of high-energy open string-string
scattering in the fixed angle regime, namely
\begin{equation}
s,-t\rightarrow\infty,t/s\approx-\sin^{2}\frac{\phi}{2}=\text{fixed (but }%
\phi\neq0\text{)}%
\end{equation}
where $s,t$ and $u$ are the Mandelstam variables and $\phi$ is the CM
scattering angle. It was shown that for the 26D open bosonic string the only
states that will survive the high-energy limit at mass level $M_{2}%
^{2}=2(N-1)$ are of the form \cite{CHLTY,PRL}%
\begin{equation}
\left\vert N,2m,q\right\rangle \equiv(\alpha_{-1}^{T})^{N-2m-2q}(\alpha
_{-1}^{L})^{2m}(\alpha_{-2}^{L})^{q}|0,k\rangle\label{2ref}%
\end{equation}
where $N,m$ and $q$ are non-negative integers and $N\geq2m+2q.$ It can be
shown that the high-energy vertex in Eq.(\ref{2ref}) are conformal invariants
up to a subleading order term in the high-energy expansion. Note that $e^{P}$
approaches to $e^{L}$ in the fixed angle regime \cite{ChanLee1}\cite{ChanLee2}%
. For simplicity, one chooses $k_{1}$, $k_{3}$ and $k_{4}$ to be tachyons. It
turns out that the high-energy fixed angle scattering amplitudes can be
calculated by using the saddle-point method. The complete ratios among the
amplitudes at each fixed mass level can be calculated to be \cite{CHLTY,PRL}%
\begin{equation}
\frac{T^{(N,2m,q)}}{T^{(N,0,0)}}=\left(  -\frac{1}{M_{2}}\right)
^{2m+q}\left(  \frac{1}{2}\right)  ^{m+q}(2m-1)!!. \label{PRL}%
\end{equation}
A calculation based on the decoupling of high-energy ZNS gave the same result
as in Eq.(\ref{PRL}).\qquad\qquad

To compare the RR amplitudes Eq.(\ref{RRamp}) with the fixed angle amplitudes
corresponding to states in Eq.(\ref{2ref}), we consider the RR amplitudes of
the following closed string states%
\begin{align}
&  |N;2m,2m^{\prime};q,q^{\prime}\rangle\nonumber\\
&  =\left(  \alpha_{-1}^{T}\right)  ^{N/2-2m-2q}\left(  \alpha_{-1}%
^{P}\right)  ^{2m}\left(  \alpha_{-2}^{P}\right)  ^{q}\otimes\left(
\tilde{\alpha}_{-1}^{T}\right)  ^{N/2-2m^{\prime}-2q^{\prime}}\left(
\tilde{\alpha}_{-1}^{P}\right)  ^{2m^{\prime}}\left(  \tilde{\alpha}_{-2}%
^{P}\right)  ^{q^{\prime}}|0,k\rangle.
\end{align}
where $m,m^{\prime},q$ and $q^{\prime}$ are non-negative integers. We can take
the following values%
\begin{align}
p_{1}  &  =N/2-2m-2q,p_{1}^{\prime}=N/2-2m^{\prime}-2q^{\prime},\\
q_{1}  &  =2m,q_{1}^{\prime}=2m^{\prime},\\
q_{2}  &  =q,q_{2}^{\prime}=q^{\prime}%
\end{align}
in Eq.(\ref{RRamp}), and include the phase factor in Eq.(\ref{phase}) to get%
\begin{align}
&  A^{(N;2m,2m^{\prime};q,q^{\prime})}=(i)^{N-q-q^{\prime}}\left(
-i\sqrt{\tilde{b}_{0}}\right)  ^{N-2\left(  m+m^{\prime}\right)  -2\left(
q+q^{\prime}\right)  }\left(  i\frac{b_{0}-1}{2M_{2}}\right)  ^{q+q^{\prime}%
}\left(  -\frac{i}{M_{2}}\right)  ^{2m+2m^{\prime}{}^{\prime}}\nonumber\\
&  \cdot B\left(  a_{0}+1,\frac{b_{0}+1-N}{2}\right)  \sum_{j=0}^{\min\left\{
2m,2m^{\prime}\right\}  }(-1)^{j}j!\binom{2m}{j}\binom{2m^{\prime}}{j}\left(
\frac{b_{0}+1-N}{2}\right)  _{j}\nonumber\\
&  \cdot U\left(  -2m-2m^{\prime}+2j,\frac{-b_{0}+N+1}{2}-2m-2m^{\prime
}+j,-\frac{b_{0}-1}{2}\right)  . \label{RRamp2}%
\end{align}
It is now easy to calculate the RR ratios for each fixed mass level
\begin{align}
\frac{A^{(N;2m,2m^{\prime};q,q^{\prime})}}{A^{(N,0,0,0,0)}}  &
=(i)^{-q-q^{\prime}}\left(  -i\frac{b_{0}-1}{2\tilde{b}_{0}M_{2}}\right)
^{q+q^{\prime}}\left(  \frac{1}{\tilde{b}_{0}M_{2}^{2}}\right)  ^{m+m^{\prime
}{}}\nonumber\\
&  \cdot\sum_{j=0}^{\min\left\{  2m,2m^{\prime}\right\}  }(-1)^{j}j!\binom
{2m}{j}\binom{2m^{\prime}}{j}\left(  \frac{b_{0}+1-N}{2}\right)
_{j}\nonumber\\
&  \cdot U\left(  -2m-2m^{\prime}+2j,\frac{-b_{0}+N+1}{2}-2m-2m^{\prime
}+j,-\frac{b_{0}-1}{2}\right)  \label{DRatio}%
\end{align}
which is a $b_{0}$-dependent function.

Before studying the fixed angle ratios for string D-particle scatterings, we
first make a pause to review previous results on \textit{string-string} scatterings.

\subsection{String-String Scatterings}

\subsubsection{Open String}

For open string-string scatterings, either the saddle-point method ($t-u$
channel only) or the decoupling of high-energy zero-norm states (ZNS) can be
used to calculate the fixed angle ratios \cite{ChanLee1,ChanLee2,
CHL,CHLTY,PRL,paperB}. It was discovered that there was an interesting link
between high-energy fixed angle amplitudes $T$ and RR amplitudes $A.$ To the
leading order in energy, the ratios among fixed angle amplitudes are $\phi
$-independent numbers, whereas the ratios among RR amplitudes are
$t$-dependent functions. However, It was discovered \cite{bosonic} that the
coefficients of the high-energy RR ratios in the leading power of $t$ can be
identified with the fixed angle ratios, namely \cite{bosonic}%
\begin{equation}
\lim_{\tilde{t}^{\prime}\rightarrow\infty}\frac{A^{(N,2m,q)}}{A^{(N,0,0,)}%
}=\left(  -\frac{1}{M_{2}}\right)  ^{2m+q}\left(  \frac{1}{2}\right)
^{m+q}(2m-1)!!=\frac{T^{(N,2m,q)}}{T^{(N,0,0)}}. \label{Ratios2}%
\end{equation}
To ensure this identification, one needs the following identity
\cite{bosonic,bosonic2,RRsusy,LYAM}
\begin{align}
&  \sum_{j=0}^{2m}(-2m)_{j}\left(  -L-\frac{\tilde{t}^{\prime}}{2}\right)
_{j}\frac{(-2/\tilde{t}^{\prime})^{j}}{j!}\nonumber\\
&  =0(-\tilde{t}^{\prime})^{0}+0(-\tilde{t}^{\prime})^{-1}+...+0(-\tilde
{t}^{\prime})^{-m+1}+\frac{(2m)!}{m!}(-\tilde{t}^{\prime})^{-m}+\mathit{O}%
\left\{  \left(  \frac{1}{\tilde{t}^{\prime}}\right)  ^{m+1}\right\}
\label{master}%
\end{align}
where $L=1-N$ and is an integer. Note that $L$ effects only the subleading
terms in $\mathit{O}\left\{  \left(  \frac{1}{\tilde{t}^{\prime}}\right)
^{m+1}\right\}  .$ Mathematically, the complete proof of Eq.(\ref{master}) for
\textit{arbitrary real values} $L$ was recently worked out in \cite{LYAM} by
using an identity of signless Stirling number of the first kind in
combinatorial theory.

\subsubsection{Open Superstring}

For all four classes \cite{susy} of high-energy fixed angle open superstring
scattering amplitudes, both the corresponding RR amplitudes and the complete
ratios of the leading (in $t$) RR amplitudes can be calculated \cite{RRsusy}.
For the fixed angle regime \cite{susy}, the complete ratios can be calculated
by the decoupling of high-energy zero norm states. It turns out that the
identification in Eq.(\ref{Ratios2}) continues to work, and $L$ is an integer
again for this case \cite{RRsusy}.

\subsubsection{Compactified Open String}

For compactified open string scatterings, both the amplitudes and the complete
ratios of leading (in $t$) RR can be calculated \cite{HLY}. For the fixed
angle regime, the complete ratios can be calculated by the decoupling of
high-energy zero norm states. The identification in Eq.(\ref{Ratios2})
continues to work. However, only scattering amplitudes corresponding to the
cases $m=0$ were calculated. The difficulties has been as following. First, it
seems that the saddle-point method is not applicable here. On the other hand,
it was shown that \cite{ChanLee1,ChanLee2,CHL} the leading order amplitudes
containing $(\alpha_{-1}^{L})^{2m}$ component will drop from energy order
$E^{4m}$ to $E^{2m}$, and one needs to calculate the complicated naive
subleading order terms in order to get the real leading order amplitude. One
encounters this difficulty even for some cases in the non-compactified string
calculation. In these cases, the method of decoupling of high-energy ZNS was adapted.

It was important to discover \cite{HLY} that the identity in Eq.(\ref{master})
for arbitrary real values $L$ can only be realized in high-energy
\textit{compactified} string scatterings. This is due to the dependence of the
value $L$ on winding momenta $K_{i}^{25}$ \cite{HLY}%
\begin{equation}
L=1-N-(K_{2}^{25})^{2}+K_{2}^{25}K_{3}^{25}. \label{L}%
\end{equation}
All other high-energy string scatterings calculated previously
\cite{bosonic,bosonic2,RRsusy} correspond to integer value of $L$ only.

\subsubsection{ Closed String}

For closed string scatterings \cite{bosonic2}, one can use the KLT formula
\cite{KLT}, which expresses the relation between tree amplitudes of closed and
two channels of open string $(\alpha_{\text{closed}}^{\prime}=4\alpha
_{\text{open}}^{\prime}=2),$ to simplify the calculations. Both ratios of
leading (in $t$) RR and fixed angle amplitudes were found to be the tensor
product of two ratios in Eq.(\ref{Ratios2}), namely \cite{bosonic2}%
\begin{align}
\lim_{\tilde{t}^{\prime}\rightarrow\infty}\frac{A_{\text{closed}}^{\left(
N;2m,2m^{^{\prime}};q,q^{^{\prime}}\right)  }}{A_{\text{closed}}^{\left(
N;0,0;0,0\right)  }}  &  =\left(  -\frac{1}{M_{2}}\right)  ^{2(m+m^{^{\prime}%
})+q+q^{^{\prime}}}\left(  \frac{1}{2}\right)  ^{m+m^{^{\prime}}%
+q+q^{^{\prime}}}(2m-1)!!(2m^{^{\prime}}-1)!!\nonumber\\
&  =\frac{T_{\text{closed}}^{\left(  N;2m,2m^{^{\prime}};q,q^{^{\prime}%
}\right)  }}{T_{\text{closed}}^{\left(  N;0,0;0,0\right)  }}. \label{RRclosed}%
\end{align}
\qquad

We now begin to discuss the RR \textit{closed string, D-particle} scatterings
considered in this paper.

\subsection{Closed String D-particle Scatterings}

\subsubsection{$m=m^{^{\prime}}=0$ Case}

In \cite{Dscatt}, the high-energy scattering amplitudes and ratios of fixed
angle closed string D-particle scatterings were calculated only for the case
$m=m^{^{\prime}}=0$. For nonzero $m$ or $m^{^{\prime}}$ cases, one encounters
similar difficulties stated in the paragraph before Eq.(\ref{L}) to calculate
the complete fixed angle amplitudes. A subset of ratios was found to be
\cite{Dscatt}%
\begin{equation}
\frac{T_{SD}^{(N,0,0,q,q^{^{\prime}})}}{T_{SD}^{(N,0,0,0,0)}}=\left(
-\frac{1}{2M_{2}}\right)  ^{q+q^{^{\prime}}}. \label{subset}%
\end{equation}
In view of the non-factorizability of Regge string D-particle scattering
amplitudes calculated in Eq.(\ref{RRamp}), one is tempted to conjecture that
the complete ratios of fixed angle closed string D-particle scatterings may
not be factorized. On the other hand, the decoupling of high-energy ZNS
implies the factorizability of the fixed angle ratios.

\subsubsection{General Case}

We can show explicitly that the leading behaviors of the inner products in
Eq.(\ref{eq1}) involving $k_{1}, k_{2}, e^{T}, e^{P}$ and $D$ are not affected
by the replacement of $e^{P}$ with $e^{L}$ if we take the limit $b_{0}
\rightarrow\infty$ after taking the Regge limit. Therefore we proceed as in
the previous works on Regge scattering. The calculation for the complete
ratios of leading (in $b_{0}$) RR \textit{closed string, D-particle}
scatterings from Eq.(\ref{DRatio}) gives%
\begin{align}
&  \lim_{b_{0}\rightarrow\infty}\frac{A_{SD}^{(N;2m,2m^{\prime};q,q^{\prime}%
)}}{A_{SD}^{(N,0,0,0,0)}}\nonumber\\
&  =(i)^{-q-q^{\prime}}\left(  -i\frac{b_{0}}{2b_{0}M_{2}}\right)
^{q+q^{\prime}}\left(  \frac{1}{{b}_{0}M_{2}^{2}}\right)  ^{m+m^{\prime}%
}\nonumber\\
&  \cdot\sum_{j=0}^{\min\left\{  2m,2m^{\prime}\right\}  }(-1)^{j}j!\binom
{2m}{j}\binom{2m^{\prime}}{j}\left(  \frac{b_{0}}{2}\right)  ^{j}\frac{\left(
2m+2m^{\prime}-2j\right)  !}{\left(  m+m^{\prime}-j\right)  !}%
2^{-2m-2m^{\prime}+2j}b_{0}^{m+m^{\prime}-j}\nonumber\\
&  =(i)^{-q-q^{\prime}}\left(  -i\frac{1}{2M_{2}}\right)  ^{q+q^{\prime}%
}\left(  \frac{1}{2M_{2}}\right)  ^{2m+2m^{\prime}}\nonumber\\
&  \cdot\sum_{j=0}^{\min\left\{  2m,2m^{\prime}\right\}  }j!\binom{2m}%
{j}\binom{2m^{\prime}}{j}\left(  -2\right)  ^{j}\frac{\left(  2m+2m^{\prime
}-2j\right)  !}{\left(  m+m^{\prime}-j\right)  !}. \label{sum}%
\end{align}
In deriving Eq.(\ref{sum}), we have made use of Eq.(\ref{Kummer}) and
Eq.(\ref{master}). Note that each term in the summation of Eq.(\ref{sum}) is
not factorized. Surprisingly, the summation in Eq.(\ref{sum}) can be
performed, and the ratios can be calculated to be%
\begin{align}
&  \lim_{b_{0}\rightarrow\infty}\frac{A_{SD}^{(N;2m,2m^{\prime};q,q^{\prime}%
)}}{A_{SD}^{(N,0,0,0,0)}}\nonumber\\
&  =(-)^{q+q^{\prime}}\left(  \frac{1}{2}\right)  ^{q+q^{\prime}%
+2m+2m^{\prime}}\left(  \frac{1}{M_{2}}\right)  ^{2m+2m^{\prime}+q+q^{\prime}%
}\nonumber\\
&  \quad\cdot\frac{2^{2m+2m^{\prime}}\pi\sec\left[  \frac{\pi}{2}\left(
2m+2m^{\prime}\right)  \right]  }{\Gamma\left(  \frac{1-2m}{2}\right)
\Gamma\left(  \frac{1-2m^{\prime}}{2}\right)  }\nonumber\\
&  =\left(  -\frac{1}{M_{2}}\right)  ^{2m+q}\left(  \frac{1}{2}\right)
^{m+q}(2m-1)!!\left(  -\frac{1}{M_{2}}\right)  ^{2m^{\prime}+q^{\prime}%
}\left(  \frac{1}{2}\right)  ^{m^{\prime}+q^{\prime}}(2m^{^{\prime}%
}-1)!!\text{ } \label{final2}%
\end{align}
which are \textit{factorized}. They are exactly the same with the ratios of
the high-energy, fixed angle closed string-string scattering amplitudes
calculated in Eq.(\ref{RRclosed}) and again consistent with the decoupling of
high-energy zero norm states \cite{ChanLee1,ChanLee2,
CHL,CHLTY,PRL,paperB,susy,Closed}. \textit{We thus conclude that the
identification in Eq.(\ref{Ratios2}) continues to work for string D-particle
scatterings.} So the complete ratios of fixed angle closed string D-particle
scatterings are%
\begin{align}
\frac{T_{SD}^{\left(  N;2m,2m^{^{\prime}};q,q^{^{\prime}}\right)  }}%
{T_{SD}^{\left(  N;0,0;0,0\right)  }}  &  =\left(  -\frac{1}{M_{2}}\right)
^{2(m+m^{^{\prime}})+q+q^{^{\prime}}}\left(  \frac{1}{2}\right)
^{m+m^{^{\prime}}+q+q^{^{\prime}}}(2m-1)!!(2m^{\prime}-1)!!\nonumber\\
&  =\lim_{b_{0}\rightarrow\infty}\frac{A_{SD}^{(N;2m,2m^{\prime};q,q^{\prime
})}}{A_{SD}^{(N,0,0,0,0)}} \label{old2}%
\end{align}
where the first equality can be deduced from the decoupling of high-energy
ZNS. Note that, for $m=m^{^{\prime}}=0,$ Eq.(\ref{old2}) reduces to
Eq.(\ref{subset}) calculated previously \cite{Dscatt}.

It is well known that the closed string-string scattering amplitudes can be
factorized into two open string-string scattering amplitudes due to the
existence of the KLT formula \cite{KLT}. On the contrary, there is no physical
picture for open string D-particle tree scattering amplitudes and thus no
factorizaion for closed string D-particle scatterings into two channels of
open string D-particle scatterings, and hence no KLT-like formula there. Here
what we really mean is: two string, two D-particle scattering in the limit of
infinite D-particle mass. This can also be seen from the nontrivial string
D-particle propagator in Eq.(\ref{DD}), which vanishes for the case of closed
string-string scattering. Thus the factorized ratios in high-energy fixed
angle regime calculated in the RR in Eq.(\ref{final2}) and Eq.(\ref{old2})
came as a surprise. However, these ratios are consistent with the decoupling
of high-energy zero norm states calculated previously \cite{ChanLee1,ChanLee2,
CHL,CHLTY,PRL,paperB,susy,Closed}. It will be interesting if one can calculate
the complete fixed angle amplitudes directly and see how the non-factorized
amplitudes can give the result of factorized ratios. We hope to pursue this
issue in the future.

\bigskip%
\setcounter{equation}{0}
\renewcommand{\theequation}{\arabic{section}.\arabic{equation}}%

\section{Conclusion}

In this paper, we study scatterings of higher spin massive closed string
states from D-particle in the Regge regime. We extract the complete infinite
ratios among high-energy scattering amplitudes of different string states in
the fixed angle regime from these Regge string scattering amplitudes. The
ratios calculated by this indirect method include a subset of ratios
calculated previously by direct fixed angle calculation \cite{Dscatt}.
Moreover, we discover that in spite of the non-factorizability of the closed
string D-particle scattering amplitudes, the complete ratios derived for the
fixed angle regime are found to be \textit{factorized}. The ratios for string
D-particle scattering amplitudes are consistent with the decoupling of
high-energy zero norm states calculated previously. \cite{ChanLee1,ChanLee2,
CHL,CHLTY,PRL,paperB,susy,Closed}.

\section{Acknowledgments}

We thank Song He, Keijiro Takahashi and Prof. C.I. Tan for helpful
discussions. This work is supported in part by the National Science Council,
50 billions project of Ministry of Education and National Center for
Theoretical Science, Taiwan.


\begin{thebibliography}{99}                                                                                               %


\bibitem {GM}D.~J.~Gross and P.~F.~Mende,
Phys.\ Lett.\ B \textbf{197}, 129 (1987);
Nucl.\ Phys.\ B \textbf{303}, 407 (1988).

\bibitem {Gross}D.~J.~Gross,
Phys.\ Rev.\ Lett.\ \textbf{60}, 1229 (1988); Phil.\ Trans.\ R. Soc. Lond.
A329, 401 (1989).

\bibitem {GrossManes}D.~J.~Gross and J.~L.~Manes,
Nucl.\ Phys.\ B \textbf{326}, 73 (1989). See section 6 for details.

\bibitem {ChanLee1}C.~T.~Chan and J.~C.~Lee,
Phys.\ Lett.\ B \textbf{611}, 193 (2005).
J.~C.~Lee,
[arXiv:hep-th/0303012].

\bibitem {ChanLee2}C.~T.~Chan and J.~C.~Lee,
Nucl.\ Phys.\ B \textbf{690}, 3 (2004).


\bibitem {CHL}C.~T.~Chan, P.~M.~Ho and J.~C.~Lee,
Nucl.\ Phys.\ B \textbf{708}, 99 (2005).


\bibitem {CHLTY}C.~T.~Chan, P.~M.~Ho, J.~C.~Lee, S.~Teraguchi and Y.~Yang,
Nucl.\ Phys.\ B \textbf{725}, 352 (2005).


\bibitem {PRL}C.~T.~Chan, P.~M.~Ho, J.~C.~Lee, S.~Teraguchi and Y.~Yang,
Phys. Rev. Lett. 96 (2006) 171601.

\bibitem {paperB}C.~T.~Chan, P.~M.~Ho, J.~C.~Lee, S.~Teraguchi and Y.~Yang,
Nucl.\ Phys.\ B \textbf{749}, 266 (2006).
\textquotedblleft Comments on the high energy limit of bosonic open string
theory,\textquotedblright\ [arXiv:hep-th/0509009].

\bibitem {susy}C.~T.~Chan, J.~C.~Lee and Y.~Yang,
Nucl.\ Phys.\ B \textbf{738}, 93 (2006).


\bibitem {Closed}C.~T.~Chan, J.~C.~Lee and Y.~Yang,
Nucl.\ Phys.\ B \textbf{749}, 280 (2006).


\bibitem {HL}Pei-Ming Ho, Xue-Yan Lin, Phys.Rev. D73 (2006) 126007.

\bibitem {ZNS1}J.~C.~Lee,
Phys.\ Lett.\ B \textbf{241}, 336 (1990); Phys.\ Rev.\ Lett.\ \textbf{64},
1636 (1990); Prog. Theor. Phys.\textbf{91}, 353 (1994). J.~C.~Lee,
Phys.\ Lett.\ B \textbf{326}, 79 (1994).


\bibitem {ZNS3}T.~D.~Chung and J.~C.~Lee,
Phys.\ Lett.\ B \textbf{350}, 22 (1995).
Z.\ Phys.\ C \textbf{75}, 555 (1997).
J.~C.~Lee,
Eur.\ Phys.\ J.\ C \textbf{1}, 739 (1998).


\bibitem {ZNS2}H.~C.~Kao and J.~C.~Lee,
Phys.\ Rev.\ D \textbf{67}, 086003 (2003).
C.~T.~Chan, J.~C.~Lee and Y.~Yang,
Phys.\ Rev.\ D \textbf{71}, 086005 (2005)


\bibitem {2DString}For a review see I.R. Klebanov and A. Pasquinucci,
hep-th/9210105 and references therein.

\bibitem {Winfinity}J.~Avan and A.~Jevicki,
Phys.\ Lett.\ B \textbf{266}, 35 (1991);
Phys.\ Lett.\ B \textbf{272}, 17 (1991). I.~R.~Klebanov and A.~M.~Polyakov,
Mod.\ Phys.\ Lett.\ A \textbf{6}, 3273 (1991).


\bibitem {Ring}E.~Witten,
Nucl.\ Phys.\ B \textbf{373}, 187 (1992).
E.~Witten and B.~Zwiebach,
Nucl.\ Phys.\ B \textbf{377}, 55 (1992).


\bibitem {Dscatt}C.~T.~Chan, J.~C.~Lee and Y.~Yang, " Scatterings of massive
string states from D-brane and their linear relations at high energies",
Nucl.Phys.B\textbf{764}, 1 (2007).

\bibitem {Decay}J.C. Lee and Y. Yang, "Linear Relations of High Energy
Absorption/Emission Amplitudes of D-brane", Phys.Lett. B646 (2007) 120, hep-th/0612059.

\bibitem {Klebanov}For a review, see A. Hashimoto and I.R. Klebanov,
"Scattering of strings from D-branes" hep-th/9611214 and references therein.
M.R. Garousi and R.C. Myers, "Superstring scattering from D-Branes" Nucl.
Phys. B475 (1996) 193, hep-th/9603194. I.R. Klebanov and L. Thorlacius, Phys.
Lett. B371,51 (1996). J.L.F. Barbon, Phys. Lett. B382, 60 (1996). C. Bachas
and B. Pioline, JHEP9912, 004 (1999). S. Hirano and Y. Kazama, Nucl. Phys.
B499, 495 (1997).

\bibitem {RR1}D. Amati, M. Ciafaloni and G. Veneziano, \textquotedblleft
Superstring Collisions at Planckian Energies,\textquotedblright Phys. Lett. B
197 (1987) 81.

\bibitem {RR2}D. Amati, M. Ciafaloni and G. Veneziano, \textquotedblleft
Classical and Quantum Gravity Effects from Planckian Energy Superstring
Collisions,\textquotedblright\ Int. J. Mod. Phys. A 3 (1988) 1615.

\bibitem {RR3}D. Amati, M. Ciafaloni and G. Veneziano, \textquotedblleft Can
Space-Time Be Probed Below The String Size?,\textquotedblright\ Phys. Lett. B
216 (1989) 41.

\bibitem {RR4}M. Soldate, \textquotedblleft Partial Wave Unitarity and Closed
String Amplitudes,\textquotedblright\ Phys. Lett. B 186 (1987) 321.

\bibitem {RR5}I. J. Muzinich and M. Soldate, \textquotedblleft High-Energy
Unitarity of Gravitation and Strings,\textquotedblright\ Phys. Rev. D 37
(1988) 359.

\bibitem {RR6}R. C. Brower, J. Polchinski, M. J. Strassler and C. I. Tan,
\textquotedblleft The pomeron and gauge / string duality,\textquotedblright\ arXiv:hep-th/0603115.

\bibitem {OA}Oleg Andreev, "More comments on the high-energy behavior of
string scattering amplitudes in warped spacetimes", Phy. Rev. D71 (2005) 066006.

\bibitem {DL}G.S. Danilov, L.N. Lipatov, "BFKL Pomeron in string models",
Nucl. Phys. B754 (2006) 187.

\bibitem {KP}M.Kachelriess, M. Plumacher, "Remarks on the high-energy behavior
of cross-sections in weak-scale string theories", hep-ph/0109184.

\bibitem {bosonic}Sheng-Lan Ko, Jen-Chi Lee and Yi Yang, "Patterns of High
energy Massive String Scatterings in the Regge regime", JHEP 0906:028,(2009);
"Kummer function and High energy String Scatterings", arXiv:0811.4502;
"Stirling number Identities and High energy String Scatterings",
arXiv:0909.3894 (published in the SLAC eConf series).

\bibitem {bosonic2}Jen-Chi Lee and Yi Yang, "Regge Closed String Scattering
and its Implication on Fixed angle Closed String Scattering", Phys.Lett.B687:84-88,2010.

\bibitem {RRsusy}S. He, J.C. Lee, K. Takahashi and Y. Yang, "Massive
Superstring Scatterings in the Regge Regime", arXiv:1001.5392. (accepted by PRD)

\bibitem {LYAM}J.C. Lee, C.H. Yan and Y. Yang, "High-energy String Scattering
Amplitudes and Signless Stirling Number Identity", arXiv: 1012.5225.

\bibitem {HLY}S. He, J.C. Lee and Y. Yang, "Exponential fall-off Behavior of
Regge Scatterings in Compactified Open String Theory", arXiv:1012.3158.

\bibitem {KLT}H. Kawai, D. Lewellen and H. Tye, "A Relation Between Tree
Amplitudes of Closed and Open Strings", Nucl.Phys.B269 (1986)1.
\end{thebibliography}
\end{document}